\documentclass[sigplan,10pt,nonacm]{acmart}
\usepackage{algorithm}
\usepackage{algpseudocode}
\usepackage{booktabs}
\usepackage{subcaption}
\usepackage[capitalize,noabbrev]{cleveref}
\usepackage{xspace}
\usepackage{amsmath}
\usepackage{xcolor}
\usepackage{tikz}
\usetikzlibrary{positioning,calc,fit,backgrounds,arrows.meta,decorations.pathreplacing}
\tikzset{
  state/.style={
    circle,
    draw=black,
    line width=0.6pt,
    fill=black!8,
    minimum size=11mm,
    inner sep=0pt,
    font=\small\itshape,
  },
  trans/.style={
    -Latex,
    line width=0.5pt,
    shorten >=0.5pt,
    shorten <=0.5pt,
  },
}

\newcommand{\evalfig}[2][\columnwidth]{%
  \IfFileExists{#2}{\includegraphics[width=#1]{#2}}{%
    \fbox{\parbox{\dimexpr#1-2\fboxsep-2\fboxrule}{\centering\vspace{4em}\textit{[figure pending: \detokenize{#2}]}\vspace{4em}}}}}

\crefname{section}{\S}{\S\S}
\Crefname{section}{\S}{\S\S}

\AtBeginDocument{%
  }

\acmConference[]{}{}{}
\acmISBN{978-1-4503-XXXX-X/2018/06}

\begin{document}

\title{\system: Efficiently Serving Diffusion LLMs with the AR Stack}

\author{Nitin Kedia}
\affiliation{%
  \institution{The University of Texas at Austin}
  \city{Austin}
  \country{United States}
}
\author{Saurabh Agarwal}
\affiliation{%
  \institution{The University of Texas at Austin}
  \city{Austin}
  \country{United States}
}
\author{Myungjin Lee}
\affiliation{%
  \institution{Cisco Research}
  \city{Bellevue}
  \country{United States}
}
\author{Aditya Akella}
\affiliation{%
  \institution{The University of Texas at Austin}
  \city{Austin}
  \country{United States}
}

\renewcommand{\shortauthors}{Kedia et al.}

\newcommand{\system}{Sangam\xspace}
\newcommand{\aditya}[1]{\textcolor{red}{AA:#1}}
\newcommand{\nitin}[1]{\textcolor{blue}{NK:#1}}
\newcommand{\saurabh}[1]{\textcolor{purple}{SA:#1}}
\newcommand{\mlee}[1]{\textcolor{violet}{ML:#1}}

\begin{abstract}
Diffusion language models (dLLMs) generate text by iteratively denoising a masked response and can commit multiple output positions per model invocation. Their bidirectional attention prevents exact autoregressive-style KV caching, since committing one position shifts the KV activations of all others.
Approximate caching techniques such as Fast-dLLM and dKV-Cache refresh KV activations repeatedly and reuse them across intervening decodes, inducing a repeated prefill/decode structure.
This makes AR serving mechanisms relevant to dLLMs, but not directly applicable. dLLM decodes are block-sized rather than token-sized, prefills recur, and bidirectional attention precludes the chunked prefill mechanism used for stall-free colocated serving. We present \system{}, a serving system for cached dLLM inference. 
\system{} introduces a deficit token-budget scheduler that admits in-flight decodes first, admits whole indivisible prefills only when the accumulated token budget allows, and carries unused budget forward. This achieves amortized stall-free scheduling.
Disaggregated serving avoids prefill-decode interference but suffers from prefill/decode resource partitioning problem. 
\system{} adopts a hybrid serving strategy, overflowing prefills onto decode workers to relieve prefill under-provisioning, and uses the same deficit-budget scheduler to protect those workers' decodes from the overflow. We show that like AR serving, dLLM serving design space is governed by prefill-decode interference and prefill/decode partitioning. Colocated serving is most effective on decode-heavy workloads, cutting mean latency by 9-20\% over hybrid execution on LLaDA-8B ShareGPT; while hybrid execution is most effective on prefill-heavy workloads, cutting mean latency by 8-20\% over colocated execution on Dream-7B arXiv. \system is available at \url{https://github.com/UT-InfraAI/sangam}.
\end{abstract}



\maketitle

\section{Introduction}

Diffusion language models (dLLMs)~\cite{sahoo2024simple, nie2026large, ye2025dream} are a recent alternative to autoregressive (AR) language models. This model class is gaining rapid traction: beyond open-weight model series such as LLaDA ~\cite{nie2026large, bie2025llada2,nie2026improved} and Dream ~\cite{ye2025dream, xie2025dream}, commercial releases including Google's DiffusionGemma ~\cite{diffusiongemma2026}, Inception's Mercury~\cite{khanna2025mercury}, and ByteDance's Seed Diffusion~\cite{song2025seed} advertise single-request generation speedups of 4-10$\times$ over comparable AR models. We study efficient serving of dLLMs under concurrent requests.

AR models generate text one token at a time from left to right. A masked-diffusion language model~\cite{sahoo2024simple} instead appends a fixed-length response region, initialized as \texttt{[MASK]} tokens, to the prompt and iteratively denoises that region. Each denoising iteration runs a Transformer~\cite{vaswani2017attention} forward pass over the prompt and response, predicts tokens for the masked positions, and commits a subset of those positions in parallel. For serving, this means that one model invocation can advance generation by multiple output tokens, whereas an AR decode invocation commits one next token. 

This execution model changes the role of key-value (KV) caching. %
In an AR model, causal attention makes previously computed KV activations stable: later tokens cannot change the KV activations of earlier tokens. A serving system can therefore compute prompt KV once during \emph{prefill} and reuse it across subsequent \emph{decode} steps. dLLMs use bidirectional attention over the full prompt-and-response sequence. When one masked position is committed, the attention inputs for other positions change, and their KV activations may also change. Exact KV reuse across denoising iterations is therefore invalid, so a naive dLLM serving system recomputes the full prompt-plus-response context at every iteration.

Recent dLLM acceleration techniques~\cite{wu2025fast, ma2026dkv, liu2026dllmcacheacceleratingdiffusionlarge, ou2025your} make this cost more manageable through approximate KV reuse. Fast-dLLM~\cite{wu2025fast} divides the fixed response region into contiguous \emph{blocks}, where a block is the subset of response positions denoised before generation advances to the next contiguous subset, and refreshes KV at block boundaries. dKV-Cache~\cite{ma2026dkv} admits KV activations of committed response tokens into the cache and periodically refreshes the cache to bound approximation error. These methods make a cached dLLM request alternate between KV-refreshing prefill phases and cache-reusing decode iterations.

This repeated prefill/decode structure makes AR serving mechanisms relevant to cached dLLMs. Existing AR serving systems provide continuous batching~\cite{yu2022orca}, colocated stall-free batching~\cite{agrawal2024taming}, disaggregated prefill/decode execution~\cite{zhong2024distserve, patel2024splitwise, hu2024inference}, and paged KV-cache management~\cite{kwon2023efficient}. These mechanisms address the right classes of control problems: iteration-level admission, prefill/decode interference, KV memory management, and placement across GPU workers. However, cached dLLMs violate several assumptions that these mechanisms rely on in AR serving. Efficient dLLM serving requires aligning these building blocks with bidirectional, cached dLLM execution.

The first mismatch is decode cost. An AR decode contributes one query token per request to an iteration. A blockwise dLLM decode contributes an entire response block while attending over the prompt and response context. Decode iteration time therefore grows much faster with batch size for dLLMs than for comparable AR models: a handful of concurrent decodes already saturate GPU compute and memory bandwidth, so admitting more decodes into a batch yields little additional throughput (\Cref{subsec:motiv-bigdecode-reprefill}). A dLLM serving system thus cannot singularly rely on large decode batches to increase throughput the way AR systems do.

The second is recurring, request-dependent prefill work. In AR serving, prefill is associated with request arrival. In cached dLLM serving, a request returns to prefill at block boundaries ~\cite{wu2025fast} or cache refresh points ~\cite{ma2026dkv}. With variable length unmasking samplers such as confidence-threshold sampling, different requests and different blocks within the same request may require different numbers of denoising iterations before the next refresh. The scheduler must therefore admit work at iteration granularity ~\cite{yu2022orca}, since one request may need to re-enter prefill while other requests continue decoding. Prefill-decode interference can arise even without new arrivals. AR colocated serving hides this interference with chunked prefill~\cite{agrawal2024taming}, splitting a prefill into pieces small enough not to stall co-batched decodes. This does not apply to dLLMs: bidirectional attention makes every token's KV activation depend on the full prompt-and-response context, so a prefill must run as one forward pass over that context and cannot be chunked.

The third mismatch is static prefill/decode partitioning. Disaggregated serving ~\cite{zhong2024distserve, patel2024splitwise, hu2024inference} removes prefill-decode interference by assigning separate GPU pools to the two phases but introduces a partitioning problem ~\cite{mitra2025beyond}. Additionally in cached dLLMs, requests cross the prefill/decode boundary repeatedly and the balance between prefill and decode work depends on the trace, load, and request progress. A fixed partition therefore creates more stranded capacity: one phase can build a queue while workers assigned to the other phase have spare capacity. 
Solutions proposed for the general prefill/decode partitioning problem are often coarse, operationally complex, and require a large number of GPUs ~\cite{qin2025mooncake,stojkovic2025dynamollm,patel2024splitwise}. 
Conditional disaggregation ~\cite{nvidia2026dynamo} handles prefill under-provisioning and is simple to implement, overflowing prefills onto decode workers when the prefill queue backs up.
This does not directly carry over to dLLMs: because decode workers are acutely sensitive to interference from injected overflow prefills, and naive conditional disaggregation for dLLMs provides no mechanism to protect those decodes from the overflowed prefill work.

We present \system{}, a serving system for cached dLLM inference. \system{} uses a central scheduler for cluster-level routing and worker-local schedulers for iteration-level batching. The central scheduler routes request-arrival prefills and cache refresh re-prefills across GPU workers using cluster-level load information, while each worker constructs microbatches locally between model invocations.

The core worker-level mechanism in \system{} is a deficit token-budget scheduler for amortized stall-free colocated serving. Each iteration has a fixed token budget. The worker admits in-flight decodes first, then admits a waiting prefill only when the budget left after decodes, plus the deficit carried over from prior iterations, can fit the whole prefill. Unused budget carries forward as deficit, preventing starvation of large indivisible prefills.
Since an indivisible prefill must enter in one iteration, per-iteration stall-free batching is unattainable; the deficit scheme instead recovers it in an amortized sense, without relying on chunked prefill ~\cite{agrawal2024taming}.
The budget remains an important scalability knob: smaller budgets protect decode iterations from prefill interference, while larger budgets admit prefill work more aggressively and can sustain higher loads when prefill queueing is the bottleneck.

Building on the deficit-budget scheduler, \system{} turns static disaggregation into a stronger design point via conditional disaggregation. We call it the \emph{hybrid} configuration which keeps a dedicated prefill pool but replaces the dedicated decode workers with deficit token budget based colocated workers. When the prefill pool becomes the bottleneck, the scheduler logically overflows prefills onto these colocated workers. For cached dLLMs this applies to both request-arrival prefills and cache refresh re-prefills, and overflowed prefills run under the same deficit token-budget policy so the operator can bound their interference with ongoing decodes.

Taken together, these mechanisms let \system{} characterize the dLLM serving design space along two fundamental problems: \textit{prefill/decode partitioning} and \textit{prefill-decode interference}, as in AR serving. Colocated and disaggregated serving sit at opposite corners of this space, the former free of partitioning but exposed to interference, the latter the reverse. We find that colocated serving is in general more suited for dLLM serving due to block-size decodes which have high GPU utilization at low batch sizes and data-dependent recurring prefills which exacerbates partitioning problems in disaggregated serving.  

\noindent\paragraph{Contributions.}
This paper's contributions are as follows:
\begin{itemize}
    \item We characterize cached dLLM serving as a repeated prefill/decode execution problem and identify the AR-serving assumptions that fail under this structure: block-sized decodes, recurring prefills, and prefill chunking (\Cref{sec:motivation}).

    \item We design a deficit token-budget scheduler which mitigates prefill-decode interference in colocated serving for dLLMs without relying on chunked prefill, which bidirectional attention precludes (\Cref{subsec:design-colocated}).

    \item We implement \system{} and evaluate colocated, static disaggregated, and hybrid serving on two open-weight dLLMs and two trace-driven workloads, showing the regimes where each mode is effective and explaining why by identifying prefill/decode partitioning and prefill/decode interference as the two fundamental factors (\Cref{subsec:eval-colocated-vs-disagg}).

    \item We demonstrate that hybrid scheduling with deficit token budgets can be used as a simple augmentation to static disaggregated deployments, keeping their performance in prefill heavy workloads and significantly improving their performance in decode heavy workloads (\Cref{subsec:eval-colocated-vs-disagg,subsec:eval-hybrid-scheduling}).
\end{itemize}

\section{Background}

We first describe the dLLM architecture and inference.

\subsection{dLLM Architecture and Inference}
\tikzset{
  tok/.style={draw,rounded corners=1.5pt,minimum height=3.6mm,
              minimum width=5.8mm,inner xsep=0.5pt,font=\small,
              line width=0.3pt,anchor=west,align=center},
  prompt/.style={tok,fill=blue!12,draw=blue!55},          %
  masked/.style={tok,fill=black!7,draw=black!35},          %
  fresh/.style={tok,fill=orange!28,draw=orange!75,line width=0.6pt}, %
  done/.style={tok,fill=teal!20,draw=teal!60},             %
  rowlab/.style={font=\itshape,anchor=east},
}

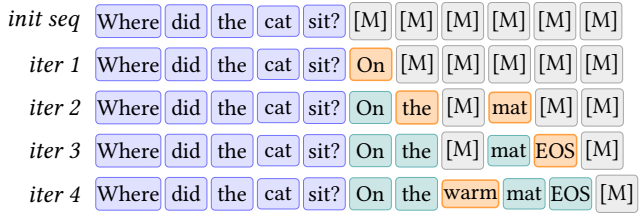
\begin{figure}[t]
\centering
\resizebox{\columnwidth}{!}{%
\begin{tikzpicture}[x=1mm,y=1mm,node distance=0.5mm]
  \def\rs{6}\def\px{10}
  \node[rowlab] at (\px-1,0) {init seq};
  \node[prompt] (q0a) at (\px,0) {Where};
  \node[prompt,right=of q0a] (q0b) {did};
  \node[prompt,right=of q0b] (q0c) {the};
  \node[prompt,right=of q0c] (q0d) {cat};
  \node[prompt,right=of q0d] (q0e) {sit?};
  \node[masked,right=of q0e] (m1) {[M]};
  \node[masked,right=of m1] (m2) {[M]};
  \node[masked,right=of m2] (m3) {[M]};
  \node[masked,right=of m3] (m4) {[M]};
  \node[masked,right=of m4] (m5) {[M]};
  \node[masked,right=of m5] (m6) {[M]};
  \node[rowlab] at (\px-1,-\rs) {iter 1};
  \node[prompt] (r1a) at (\px,-\rs) {Where};
  \node[prompt,right=of r1a] (r1b) {did};
  \node[prompt,right=of r1b] (r1c) {the};
  \node[prompt,right=of r1c] (r1d) {cat};
  \node[prompt,right=of r1d] (r1e) {sit?};
  \node[fresh,right=of r1e] (s1) {On};
  \node[masked,right=of s1] (s2) {[M]};
  \node[masked,right=of s2] (s3) {[M]};
  \node[masked,right=of s3] (s4) {[M]};
  \node[masked,right=of s4] (s5) {[M]};
  \node[masked,right=of s5] (s6) {[M]};
  \node[rowlab] at (\px-1,-2*\rs) {iter 2};
  \node[prompt] (r2a) at (\px,-2*\rs) {Where};
  \node[prompt,right=of r2a] (r2b) {did};
  \node[prompt,right=of r2b] (r2c) {the};
  \node[prompt,right=of r2c] (r2d) {cat};
  \node[prompt,right=of r2d] (r2e) {sit?};
  \node[done,right=of r2e] (t1) {On};
  \node[fresh,right=of t1] (t2) {the};
  \node[masked,right=of t2] (t3) {[M]};
  \node[fresh,right=of t3] (t4) {mat};
  \node[masked,right=of t4] (t5) {[M]};
  \node[masked,right=of t5] (t6) {[M]};
  \node[rowlab] at (\px-1,-3*\rs) {iter 3};
  \node[prompt] (r3a) at (\px,-3*\rs) {Where};
  \node[prompt,right=of r3a] (r3b) {did};
  \node[prompt,right=of r3b] (r3c) {the};
  \node[prompt,right=of r3c] (r3d) {cat};
  \node[prompt,right=of r3d] (r3e) {sit?};
  \node[done,right=of r3e] (u1) {On};
  \node[done,right=of u1] (u2) {the};
  \node[masked,right=of u2] (u3) {[M]};
  \node[done,right=of u3] (u4) {mat};
  \node[fresh,right=of u4] (u5) {EOS};
  \node[masked,right=of u5] (u6) {[M]};
  \node[rowlab] at (\px-1,-4*\rs) {iter 4};
  \node[prompt] (r4a) at (\px,-4*\rs) {Where};
  \node[prompt,right=of r4a] (r4b) {did};
  \node[prompt,right=of r4b] (r4c) {the};
  \node[prompt,right=of r4c] (r4d) {cat};
  \node[prompt,right=of r4d] (r4e) {sit?};
  \node[done,right=of r4e] (v1) {On};
  \node[done,right=of v1] (v2) {the};
  \node[fresh,right=of v2] (v3) {warm};
  \node[done,right=of v3] (v4) {mat};
  \node[done,right=of v4] (v5) {EOS};
  \node[masked,right=of v5] (v6) {[M]};
\end{tikzpicture}%
}

\caption{Diffusion LLM inference for the prompt "Where did the cat sit?" and response "On the warm mat". 
A dLLM appends a fixed-length block of \textsc{mask} tokens (shown as \texttt{[M]}) to the prompt and unmasks a subset of positions in parallel and out of order each iteration. Generation stops once \textsc{eos} is committed (iter 3) and every position before it is unmasked (iter 4); positions after \textsc{eos} are ignored.}
\label{fig:inference}
\end{figure}
dLLMs are masked-diffusion sequence models ~\cite{sahoo2024simple,shi2024simplified,austin2021structured} that treat generation as iterative denoising rather than next-token prediction. Architecturally, they reuse the Transformer decoder blocks ~\cite{vaswani2017attention}. However, the causal mask is dropped so tokens attend bidirectionally over the full prompt-and-response.

A request supplies a prompt $p$ and a maximum generation length $L$. The response is initialized to $L$ \texttt{[MASK]} tokens which are progressively unmasked using several denoising steps. Each step runs a forward pass with the full sequence of $[p; L]$ tokens, predicts a token distribution at every masked position in parallel, and \emph{unmasks} a subset of those predictions while keeping the rest masked. Unlike an AR model, which commits one new token left to right per iteration, a dLLM appends a fixed-length block of \texttt{[MASK]} tokens to the prompt and unmasks positions in parallel and out of order until the response is complete (\Cref{fig:inference}). Each such denoising step is structurally an AR-style \emph{prefill}: it is typically compute-bound. KV caching, a standard strategy in AR serving, does not directly apply: because attention in dLLM is bidirectional, unmasking a single position in one step shifts the keys and values of every other token in the next step.

\subsection{Caching in dLLM Inference}
\newlength{\figh}\setlength{\figh}{22mm}
 
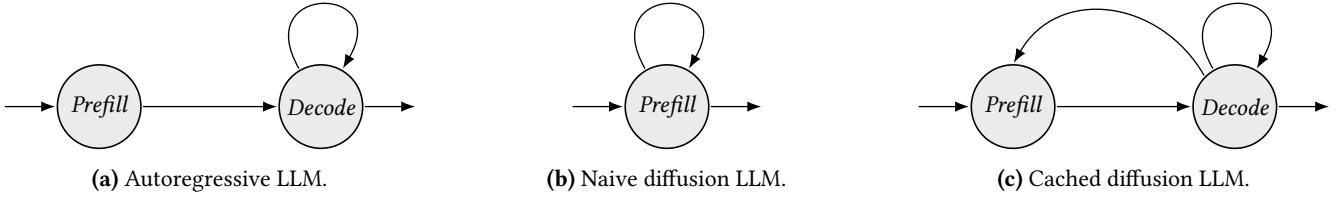
\begin{figure*}[t]
  \centering
  \begin{subfigure}[b]{0.32\textwidth}
    \centering
    \resizebox{!}{\figh}{\begin{tikzpicture}[node distance=18mm]
  \node[state] (prefill) {Prefill};
  \node[state, right=of prefill] (decode) {Decode};

  \coordinate (in)  at ($(prefill.west) + (-7mm,0)$);
  \coordinate (out) at ($(decode.east)  + ( 7mm,0)$);

  \draw[trans] (in)      -- (prefill);
  \draw[trans] (prefill) -- (decode);
  \draw[trans] (decode)  edge[loop above, looseness=6, in=60, out=120] (decode);
  \draw[trans] (decode)  -- (out);
\end{tikzpicture}\unskip}
    \caption{Autoregressive LLM.}
    \label{fig:state-ar}
  \end{subfigure}\hfill
  \begin{subfigure}[b]{0.32\textwidth}
    \centering
    \resizebox{!}{\figh}{\begin{tikzpicture}[node distance=18mm]
  \node[state] (prefill) {Prefill};

  \coordinate (in)  at ($(prefill.west) + (-7mm,0)$);
  \coordinate (out) at ($(prefill.east) + ( 7mm,0)$);

  \draw[trans] (in)      -- (prefill);
  \draw[trans] (prefill) edge[loop above, looseness=6, in=60, out=120] (prefill);
  \draw[trans] (prefill) -- (out);
\end{tikzpicture}\unskip}
    \caption{Naive diffusion LLM.}
    \label{fig:state-naive-dllm}
  \end{subfigure}\hfill
  \begin{subfigure}[b]{0.32\textwidth}
    \centering
    \resizebox{!}{\figh}{\begin{tikzpicture}[node distance=18mm]
  \node[state] (prefill) {Prefill};
  \node[state, right=of prefill] (decode) {Decode};

  \coordinate (in)  at ($(prefill.west) + (-7mm,0)$);
  \coordinate (out) at ($(decode.east)  + ( 7mm,0)$);

  \draw[trans] (in)      -- (prefill);
  \draw[trans] (prefill) -- (decode);
  \draw[trans] (decode)  edge[loop above, looseness=6, in=60, out=120] (decode);
  \draw[trans] (decode)  -- (out);

  \draw[trans] (decode.north west) to[out=120, in=60, looseness=1.2] (prefill.north);
\end{tikzpicture}\unskip}
    \caption{Cached diffusion LLM.}
    \label{fig:state-cached-dllm}
  \end{subfigure}
  \caption{Execution state machines for three LLM serving regimes.
    The autoregressive baseline (\subref{fig:state-ar}) prefills once
    and then decodes one token at a time. A naive diffusion LLM
    (\subref{fig:state-naive-dllm}) re-runs prefill at every denoising
    step. Adding a KV cache to the diffusion model
    (\subref{fig:state-cached-dllm}) reintroduces a decode state but
    still requires an occasional return to prefill when the cache
    must be refreshed.}
  \label{fig:state-machines}
\end{figure*}

Recent dLLM serving optimizations, such as Fast-dLLM~\cite{wu2025fast} and dKV-Cache~\cite{ma2026dkv}, observe that KV activations remain sufficiently stable across adjacent denoising steps to permit approximately lossless caching. \Cref{fig:state-machines} summarizes the execution state machines these regimes induce. This caching avoids the naive dLLM approach of executing a full-sequence prefill at every step (\Cref{fig:state-naive-dllm}). Instead, these systems reintroduce a decode state by reusing KV caches over short windows, occasionally forcing a return to the prefill state to refresh the cache and bound activation drift (\Cref{fig:state-cached-dllm}).

Fast-dLLM achieves this via blockwise unmasking, which trades any-order expressivity for cacheability by constraining generation to advance autoregressively across token blocks. The cache is fully recomputed at each block boundary and reused for all internal denoising steps. In contrast, dKV-Cache ~\cite{ma2026dkv} preserves any-order diffusion by caching at the per-token level. Using its \emph{dKV-Cache-PD} variant, the prompt KV is computed once and fixed, while the KV of decoded response tokens is intermittently refreshed to prevent drift.

Consequently, both schedules transform the dLLM request lifecycle into a cyclic prefill-decodes-prefill pattern, fundamentally diverging from the standard AR model of a single prefill followed by many lightweight decodes (\Cref{fig:state-ar}). These dLLM "decodes" are vastly more computationally intensive than their AR counterparts. Because a dLLM decode involves a multi-token forward pass attending to nearly the entire sequence length, its per-step cost more closely resembles a small AR chunked prefill ~\cite{agrawal2024taming}. Caching makes dLLM execution resemble AR serving closely enough to inherit much of its optimization literature, but the resemblance is not exact. The cyclic recurrence of heavy decodes and forced prefills, together with bidirectional attention, breaks assumptions embedded in AR infrastructure, so importing these techniques demands substantial system-level adaptation.

\subsection{LLM Serving}\label{subsec:background-llm-serving}
Before introducing challenges for serving dLLMs using AR serving stack, we briefly explain the key techniques. 

\noindent\textbf{{Continuous Batching.}} 
Orca~\cite{yu2022orca} introduced \emph{continuous batching}, which dynamically schedules requests per model iteration to minimize queue delay and prevent lifetime batch locking. However, co-scheduling newly arrived prefills with in-flight decodes introduces \emph{prefill-decode interference}. Prefills are compute-bound operations over large prompt sequences, whereas decodes are memory-bound single-token operations. Mixing them forces the shared iteration time to scale with the incoming prompt length, meaning a newly admitted prefill causes multi-second \emph{generation stalls} \cite{agrawal2024taming} for all concurrently decoding requests.

\noindent\textbf{Disaggregated Serving.} To solve this interference problem,  one solution is disaggregated serving ~\cite{zhong2024distserve, patel2024splitwise, hu2024inference} which takes the \emph{physical} route of processing prefill and decode onto disjoint GPU pools so the two phases never share an iteration. The KV state produced by a prefill worker is transferred to a decode worker before generation continues. This eliminates phase-mix scheduling entirely, trading interference for partitioning, placement, and KV migration cost. 

\noindent\textbf{Chunked Prefill}.
Another solution to the interference problem is using chunked prefill combined with stall-free batching. 
Sarathi-Serve~\cite{agrawal2024taming} 
introduces these two techniques. It takes the \emph{colocated} route of mitigating the stall while keeping prefill and decode on the same GPU. It introduces two ideas: \emph{chunked prefills} that split a long prompt into pieces so no single iteration carries the full prefill cost, and \emph{stall-free batching} that admits ongoing decodes first and then tops up the iteration with prefill chunks under a per iteration token budget $\tau$. The token budget is the throughput-versus-latency knob, and $\tau$ is sized so that the resulting mixed iteration balances both Time Between Tokens (TBT) SLOs and saturating the GPU compute.

\section{Motivation}\label{sec:motivation}

We discuss properties of LLM serving that need to be modified for serving dLLMs. 

\subsection{Big \textit{decodes}}\label{subsec:motiv-bigdecode-reprefill}

\begin{figure}[t]
    \centering
    \begin{subfigure}[t]{0.49\columnwidth}
        \centering
        \evalfig[\linewidth]{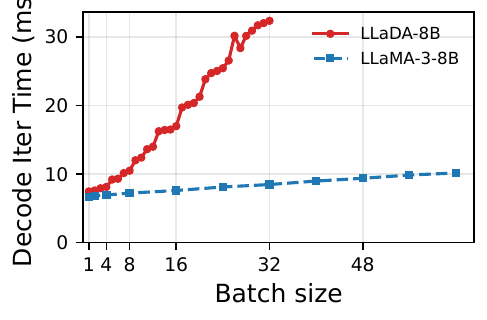}
        \caption{LLaDA-8B vs Llama-3-8B.}
        \label{fig:motiv-decode-itertime-llada-llama}
    \end{subfigure}\hfill
    \begin{subfigure}[t]{0.49\columnwidth}
        \centering
        \evalfig[\linewidth]{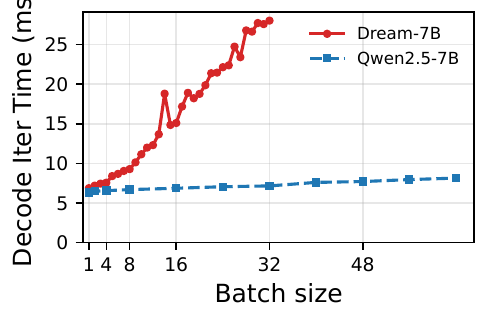}
        \caption{Dream-7B vs Qwen2.5-7B.}
        \label{fig:motiv-decode-itertime-dream-qwen}
    \end{subfigure}
    \caption{Decode iteration time vs batch size for two dLLMs (LLaDA-8B, Dream-7B) and their AR counterparts (Llama-3-8B, Qwen2.5-7B) at sequence length 1024. dLLM iteration time climbs steeply, so batching past a few requests stops improving throughput. AR iteration time grows only gently out to batch 64, so AR keeps benefiting from batching.}
    \label{fig:motiv-decode-itertime}
\end{figure}

\Cref{fig:motiv-decode-itertime} plots decode iteration time against batch size for two dLLMs (LLaDA-8B, Dream-7B) and their AR counterparts\footnote{The architecture of LLaDA-8B and Dream-7B have been derived from LLaMA-3-8B and Qwen-2.5-7B respectively.} (Llama-3-8B ~\cite{grattafiori2024llama}, Qwen2.5-7B ~\cite{qwen2025qwen25technicalreport}, measured on SGLang ~\cite{zheng2024sglang}). For both dLLMs, iteration time grows several-fold across the measured range, while for the AR models it grows only modestly. Decode batching is therefore significantly less impactful for high-throughput low-latency dLLM serving compared to AR LLMs.

The parallel unmasking mechanics of dLLMs impose a fundamentally heavier compute load during batch execution compared to traditional autoregressive generation. The primary throughput gains of decode batching stem from amortizing the memory-bandwidth bound loading of weights in MLP operations. In AR decode, each request appends only a single query token, so the marginal MLP cost per request is negligible and many requests can be packed before the MLP becomes compute-bound. Attention remains memory-bandwidth-bound and scales per request in both regimes. The difference is on the MLP path. Under block-based approximate caching such as Fast-dLLM, each active dLLM request contributes its entire current unmasking block (32 tokens by default) to the microbatch at every step. The MLP operations that batching is meant to amortize now receive a full block per request rather than a single token, so the per-request MLP cost is roughly block\_size× larger.

Consequently, these individually \textit{big} decodes restrict the throughput gains typically achieved via large decode batches. Unlike standard AR engines that rely on massive batches to maximize hardware utilization, the block-wise unmasking of dLLMs saturates GPU compute and memory bandwidth with only a few concurrent requests.

\noindent\textit{\textbf{Takeaway 1}: Each decode contributes a full block of compute, so decode batches saturate GPUs at low batch sizes. Because of this, dLLM inference engines cannot singularly rely on creating large batches to drive throughput improvements.
}

\subsection{Recurring Prefills and Variable Iterations Across Requests}\label{subsec:motiv-reprefills}

\begin{figure}[t]
    \centering
    \begin{subfigure}[t]{0.49\columnwidth}
        \centering
        \evalfig[\linewidth]{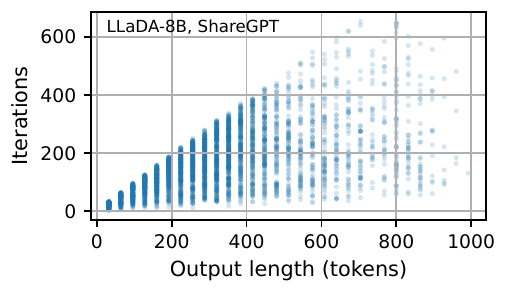}
        \caption{LLaDA-8B, ShareGPT.}
        \label{fig:motiv-iters-vs-len-llada-sharegpt}
    \end{subfigure}\hfill
    \begin{subfigure}[t]{0.49\columnwidth}
        \centering
        \evalfig[\linewidth]{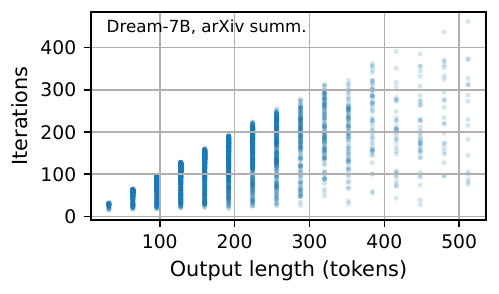}
        \caption{Dream-7B, arXiv summ.}
        \label{fig:motiv-iters-vs-len-dream-arxiv}
    \end{subfigure}
    \caption{Iterations (forward passes) versus output length, one point per request. Because the number of positions unmasked per iteration is data-dependent, requests that emit the same number of tokens still take widely different numbers of iterations to finish, visible as the vertical spread at any fixed output length.}
    \label{fig:motiv-iters-vs-len}
\end{figure}

Under blockwise denoising, every block ends with a re-prefill, but the iteration on which that boundary arrives is data-dependent. Recent sampling strategies ~\cite{ghazvininejad2019mask, wu2025fast,wei2026acceleratingdiffusionlargelanguage} unmask a \emph{variable} number of tokens per iteration, so the step count needed to unmask a block is not fixed. Confidence-threshold based sampling unmasks every masked position whose top-1 probability exceeds a threshold and lets the per-step yield vary with how confident the model happens to be.
Prompt-dependent complexity  compounds the variability: under the same sampling rule, easy prompts can resolve a block in a handful of iterations while harder ones need many more. We show this empirically. In \Cref{fig:motiv-iters-vs-len}, each point is one request's total iteration (forward-pass) count plotted against its output length, for LLaDA-8B on ShareGPT and Dream-7B on arXiv summarization traces (~\Cref{tab:eval-traces}). The block size is fixed, so output length determines the number of blocks, and all requests at a given output length denoise the same number of blocks. Yet at any fixed output length the points still spread vertically over a wide range, so requests with identical block counts take very different total iteration counts. The spread can only come from per-block variation: individual blocks unmask in different numbers of iterations, and two requests even with the same number of blocks reach their block boundaries on different iterations.

The first consequence is for batch scheduling. If every co-batched request crossed its block boundaries on the same iteration, the server could switch the whole batch between prefill and decode in lockstep.
The variability above rules this out: requests batched together reach their block boundaries on different iterations, so there is no single iteration at which the batch as a whole changes phase. Each request must instead be advanced on its own boundary schedule: on every iteration the server decides, per request, whether it is decoding within a block, re-prefilling, or finished. Therefore continuous batching that Orca ~\cite{yu2022orca} introduced for AR serving is needed: the scheduler decides on each iteration which requests run, and a request that finishes its current block can immediately submit its re-prefill while the rest of the batch keeps decoding.

The second consequence is for prefill-decode interference. A freshly arrived prompt prefill, or any in-flight request refreshing its KV cache (re-prefill), stalls the decodes of every other co-batched request. AR produces prefill work only at request arrival, so a steady-state decode batch sees interference only from new request arrivals. In dLLMs every in-flight request is also a recurring prefill source, since each block boundary re-injects a prefill. This increases prefill-decode interference. AR serving uses two techniques to mitigate this interference: Colocated serving with chunked prefills (Sarathi-Serve \cite{agrawal2024taming}) and disaggregated serving ~\cite{zhong2024distserve, patel2024splitwise, hu2024inference}, as discussed in ~\Cref{subsec:background-llm-serving}. Sarathi-Serve only adds partial prefills to ongoing decode batches to limit the interference caused by prefills. dLLMs do not support prefill chunking. This is because of bidirectional attention, where prompt tokens attend to each other (and the placeholder response tokens), affecting each other's KV values. So the AR KV-Cache invariant that KV values of prior tokens are not affected by subsequent tokens doesn't hold, and thus prefills cannot be processed in chunks. We discuss disaggregated serving in the next subsection ~\Cref{subsec:motiv-static-partition}.

\noindent\textit{\textbf{Takeaway 2}: Variable-token samplers make block boundaries a per-request, data-dependent event, so continuous batching at the iteration boundary is required for dLLM serving. Furthermore, prefill-decode interference occurs far more often than in AR because every in-flight request becomes a recurring prefill source. But dLLMs don't support prefill chunking, a critical mechanism to mitigate this interference.}

\subsection{Static prefill-decode worker partitioning is fragile}\label{subsec:motiv-static-partition}

\begin{figure}[t]
    \centering
    \includegraphics[width=0.6\columnwidth]{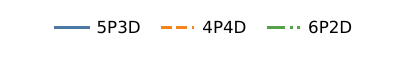}\\[-0.25em]
    \begin{subfigure}[t]{0.48\columnwidth}
        \centering\evalfig[\linewidth]{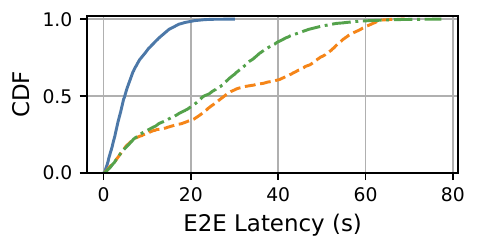}
        \caption{e2e latency}\label{fig:motiv-sp-sharegpt-e2e}
    \end{subfigure}\hfill
    \begin{subfigure}[t]{0.48\columnwidth}
        \centering\evalfig[\linewidth]{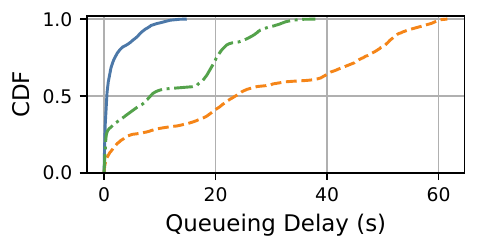}
        \caption{scheduling delay}\label{fig:motiv-sp-sharegpt-sched}
    \end{subfigure}
    \caption{LLaDA-8B on a single 8-GPU H100 node under three static disaggregated splits (4P4D, 5P3D, 6P2D) on the ShareGPT trace at QPS 6.5. 4P4D and 6P2D both inflate e2e latency, but for different reasons: under 4P4D the scheduling delay accrues at the prefill workers, while under 6P2D it accrues at the decode workers.}
    \label{fig:motiv-static-partition}
\end{figure}

Disaggregation is appealing for dLLM serving because it removes prefill-decode interference by design. 
This benefit is not free: it presupposes a prefill:decode partition whose optimum depends on the workload mix and SLO targets ~\cite{mitra2025beyond, qin2025mooncake,stojkovic2025dynamollm,patel2024splitwise}. Systems that adjust this partition online exist, but they rebalance coarsely, leave GPUs underutilized between adjustments, and add operational complexity (we return to these in detail below).
We characterize how brittle static disaggregated serving is in \Cref{fig:motiv-static-partition} by serving 8 LLaDA-8B replicas on an 8-GPU node and sweeping prefill:decode worker splits (4P4D, 5P3D, 6P2D) on the ShareGPT trace at QPS\, 6.5. 5P3D has the lowest end-to-end time. However both 4P4D and 6P2D expose two distinct failure modes that inflate e2e latency (\Cref{fig:motiv-sp-sharegpt-e2e}) through queueing delay (\Cref{fig:motiv-sp-sharegpt-sched}), but on opposite sides of the pipeline. At 4P4D prefill capacity is insufficient: queueing delay accrues at the prefill workers, reaching a median near 24\,s and lifting the e2e median to 28\,s. Reallocating one GPU to prefill (6P2D) relieves the prefill workers, but the now-smaller decode pool exhausts its memory capacity and queueing delay instead accrues at the decode workers (median ${\sim}8$\,s), still leaving an e2e median of 23\,s. Only the balanced 5P3D controls both, with scheduling delay near zero and an e2e median under 5\,s. Finer-grained partitions are possible with more GPUs, but the optimal allocation still shifts with request load and composition.

Adjusting the split online is itself a hard problem. Mooncake ~\cite{qin2025mooncake}, a production system, rebalances its prefill and decode pools with a controller that follows a daily machine tide, adding capacity to the overloaded side and reclaiming it from the other, but it does so only on a coarse, diurnal timescale by scaling cluster-level capacity, so a pool can stay over- or under-provisioned between adjustments and the scheme leans on spare GPUs. This approach is available to large clusters but not to smaller deployments which have few workers, coarse ratios, and no idle headroom to spare.
A finer-grained alternative lets decode workers absorb prefill work when the prefill side is overloaded: Splitwise~\cite{patel2024splitwise} morphs machines between roles through a mixed pool, and the conditional disaggregation in Dynamo~\cite{nvidia2026dynamo} processes a request's prefill on the decode worker when the remote prefill queue is backed up. 
However, overflow prefills on otherwise-decode workers are especially sensitive to interference. In disaggregated serving, only the decode pool, a fraction of the workers, carries all decode work, concentrating it on few workers already near GPU saturation (§3.1). Therefore an injected prefill stalls their in-flight decodes rather than filling slack. A mechanism is therefore needed that protects these decode workers while still sparing some capacity to an underprovisioned prefill side.

There is also a prior question of which regime to use at all. For AR serving, it is well known that disaggregation suits prefill-heavy workloads and colocation decode-heavy workloads ~\cite{mitra2025beyond}. In dLLMs, each decode is heavier (\Cref{subsec:motiv-bigdecode-reprefill}) and the prefill work per request grows because every block boundary triggers a re-prefill (\Cref{subsec:motiv-reprefills}). Therefore, it is not clear which regime is more suited for dLLM serving which we evaluate in \Cref{subsec:eval-colocated-vs-disagg}. 

\noindent\textit{\textbf{Takeaway 3}: No static prefill:decode split is jointly optimal: prefill-biased splits starve decode of KV Cache while decode-biased splits queue prefills. Existing online remedies are either coarse or operationally complex, and still leave interference-sensitive decode workers exposed. We instead pursue hybrid scheduling (\Cref{subsec:design-hybrid}): a fast-reacting, operationally simple mechanism that overflows a saturated prefill pool into protected decode-role workers.}

\section{\system Design}\label{sec:design}
Based on the characteristics of dLLM's discussed in previous section, we introduce \system. First, we discuss the core building blocks of \system (\S\ref{subsec:design-colocated}, \S\ref{subsec:design-hybrid}) before finally discussing the architecture (\S\ref{subsec:architecture}).

\subsection{Colocated serving with \textit{deficit} token budgets}\label{subsec:design-colocated}

\definecolor{decodeFill}{HTML}{2C7FB8}
\definecolor{decodeEdge}{HTML}{1A4E73}
\definecolor{prefillFill}{HTML}{E6792B}
\definecolor{prefillEdge}{HTML}{A8501A}
\definecolor{waitFill}{HTML}{F2C9A0}
\definecolor{waitEdge}{HTML}{C98B4F}
\definecolor{deficitFill}{HTML}{74C476}
\definecolor{deficitEdge}{HTML}{2E7D32}
\definecolor{budgetLine}{HTML}{555555}

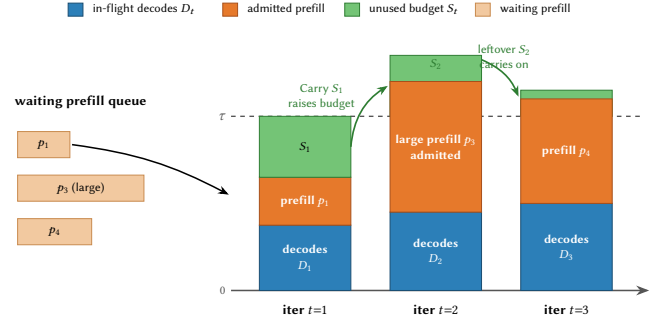
\begin{figure}[t]
\centering
\resizebox{\columnwidth}{!}{%
\begin{tikzpicture}[font=\sffamily\small]

\def\hw{1.05}
\def\tauY{4.0}

\coordinate (c1) at (0,0);
\coordinate (c2) at (3.0,0);
\coordinate (c3) at (6.0,0);

\newcommand{\blk}[6]{%
  \fill[#4,draw=#5,line width=0.6pt] (#1-\hw, #2) rectangle (#1+\hw, #2+#3);
  \node[text=white,font=\sffamily\footnotesize\bfseries,align=center] at (#1, #2+#3/2) {#6};
}
\newcommand{\blkdark}[6]{%
  \fill[#4,draw=#5,line width=0.6pt] (#1-\hw, #2) rectangle (#1+\hw, #2+#3);
  \node[font=\sffamily\footnotesize\bfseries,align=center] at (#1, #2+#3/2) {#6};
}

\def\Lx{-1.7}
\def\Rx{7.7}
\def\legShift{-5.44}

\draw[budgetLine,dashed,line width=0.9pt] (\Lx,\tauY) -- (\Rx,\tauY);
\node[budgetLine,anchor=east,font=\sffamily\bfseries] at (\Lx,\tauY) {$\tau$};
\draw[budgetLine,-{Stealth[length=3mm]},line width=1pt] (\Lx,0) -- (\Rx+0.2,0);
\node[budgetLine,anchor=east,font=\sffamily\footnotesize] at (\Lx,0) {$0$};

\blk{0}{0}{1.5}{decodeFill}{decodeEdge}{decodes\\$D_1$}
\blk{0}{1.5}{1.1}{prefillFill}{prefillEdge}{prefill $p_1$}
\blkdark{0}{2.6}{1.4}{deficitFill}{deficitEdge}{$S_1$}

\fill[deficitFill,draw=deficitEdge,line width=0.6pt] (3.0-\hw,\tauY) rectangle (3.0+\hw,\tauY+1.4);
\node[deficitEdge,font=\sffamily\footnotesize\bfseries] at (3.0,{\tauY+1.2}) {$S_2$};
\blk{3.0}{0}{1.8}{decodeFill}{decodeEdge}{decodes\\$D_2$}
\blk{3.0}{1.8}{3.0}{prefillFill}{prefillEdge}{large prefill $p_3$\\admitted}

\fill[deficitFill,draw=deficitEdge,line width=0.6pt] (6.0-\hw,\tauY) rectangle (6.0+\hw,\tauY+0.6);
\blk{6.0}{0}{2.0}{decodeFill}{decodeEdge}{decodes\\$D_3$}
\blk{6.0}{2.0}{2.4}{prefillFill}{prefillEdge}{prefill $p_4$}

\node[anchor=north,font=\sffamily\small\bfseries] at (0,-0.2) {iter $t{=}1$};
\node[anchor=north,font=\sffamily\small\bfseries] at (3.0,-0.2) {iter $t{=}2$};
\node[anchor=north,font=\sffamily\small\bfseries] at (6.0,-0.2) {iter $t{=}3$};

\draw[deficitEdge,-{Stealth[length=2.5mm]},line width=1.1pt]
  (0+\hw,3.3) to[bend left=26] node[pos=0.45,above left=-1pt,font=\sffamily\footnotesize,deficitEdge,align=center]{Carry $S_1$\\raises budget} (3.0-\hw-0.05,\tauY+0.7);
\draw[deficitEdge,-{Stealth[length=2.5mm]},line width=1.1pt]
  (3.0+\hw,\tauY+1.3) to[bend left=22] node[pos=0.5,above,font=\sffamily\footnotesize,deficitEdge,align=center]{leftover $S_2$\\carries on} (6.0-\hw-0.05,\tauY+0.45);

\node[anchor=south,font=\sffamily\small\bfseries] at (-5.15,4.05) {waiting prefill queue};
\fill[waitFill,draw=waitEdge,line width=0.6pt] (-6.6,3.05) rectangle (-5.4,3.65);
\node[font=\sffamily\footnotesize] at (-6.0,3.35) {$p_1$};
\fill[waitFill,draw=waitEdge,line width=0.6pt] (-6.6,2.05) rectangle (-3.7,2.65);
\node[font=\sffamily\footnotesize] at (-5.15,2.35) {$p_3$ (large)};
\fill[waitFill,draw=waitEdge,line width=0.6pt] (-6.6,1.05) rectangle (-4.9,1.65);
\node[font=\sffamily\footnotesize] at (-5.75,1.35) {$p_4$};
\draw[-{Stealth[length=2.5mm]},line width=1pt] (-5.35,3.35) to[bend left=10] (-1.75,2.2);

\begin{scope}[shift={(\legShift,6.3)}]
  \def\sw{0.34}
  \fill[decodeFill,draw=decodeEdge] (0,0) rectangle (\sw,0.30);
  \node[anchor=west,font=\sffamily\footnotesize] at (\sw+0.1,0.15){in-flight decodes $D_t$};
  \fill[prefillFill,draw=prefillEdge] (3.55,0) rectangle (3.55+\sw,0.30);
  \node[anchor=west,font=\sffamily\footnotesize] at (3.55+\sw+0.1,0.15){admitted prefill};
  \fill[deficitFill,draw=deficitEdge] (6.35,0) rectangle (6.35+\sw,0.30);
  \node[anchor=west,font=\sffamily\footnotesize] at (6.35+\sw+0.1,0.15){unused budget $S_t$};
  \fill[waitFill,draw=waitEdge] (9.35,0) rectangle (9.35+\sw,0.30);
  \node[anchor=west,font=\sffamily\footnotesize] at (9.35+\sw+0.1,0.15){waiting prefill};
\end{scope}

\end{tikzpicture}%
}
\caption{Deficit token-budget scheduling over three iterations. Each iteration first fills the per-iteration budget $\tau$ with in-flight decodes $D_t$, then admits waiting prefills greedily in request arrival order while they fit in the remaining budget $R_t = \tau - D_t + S_{t-1}$. Any unspent budget $S_t$ is carried forward, raising the effective ceiling in the next iteration so that a large prefill (e.g.\ $p_3$) can be admitted without exceeding the amortized per-iteration budget $\tau$.}
\label{fig:deficit-illustration}
\end{figure}
\begin{algorithm}[t]
\caption{\textit{Deficit token-budget scheduling}}
\label{alg:deficit}
\begin{algorithmic}[1]
\renewcommand{\algorithmicrequire}{\textbf{Input:}}
\renewcommand{\algorithmicensure}{\textbf{State:}}
\Require Token Budget $\tau$;\,\textit{deficit} from the previous iteration
\Ensure In-flight \textit{decodes}; queue of waiting \textit{prefills}
\If{\textit{prefills} is empty}
    \State \textit{deficit} $\gets 0$
    \State \Return \textit{decodes} \Comment{decode-only iteration}
\EndIf
\State \textit{budget} $\gets \tau - \Call{Tokens}{\textit{decodes}} + \textit{deficit}$
\State \textit{microbatch} $\gets \textit{decodes}$ \Comment{decodes are never displaced}
\While{\textit{prefills} is not empty}
    \State $p \gets$ first request in \textit{prefills}
    \State \textit{idle} $\gets$ (\textit{microbatch} is empty)
    \If{$\Call{Tokens}{p} \le \textit{budget}$ \textbf{ or } \textit{idle}}
        \State move $p$ from \textit{prefills} into \textit{microbatch}
        \State \textit{budget} $\gets \textit{budget} - \Call{Tokens}{p}$
    \Else
        \State \textbf{break} \Comment{$p$ waits; prefills cannot be split}
    \EndIf
\EndWhile
\State \textit{deficit} $\gets \max(\textit{budget},\ 0)$ \Comment{carry leftover}
\State \Return \textit{microbatch}
\end{algorithmic}
\end{algorithm}

To enable amortized stall-free batching without prefill chunking, we introduce  a \emph{deficit token budget} inspired by Deficit Round-Robin~\cite{shreedhar1995efficient}, illustrated in \Cref{fig:deficit-illustration}. Each iteration carries a token budget $\tau$. The scheduler first seats every in-flight decode, which is never displaced, so the budget left for fresh prefill work is $\textit{budget} = \tau - \textsc{Tokens}(\textit{decodes}) + \textit{deficit}$, where $\textit{deficit}$ is the unspent budget carried forward from the previous iteration. 
Waiting prefills (both new prefills and block-boundary re-prefills) are admitted greedily in request arrival-time order, not strict FIFO, so a re-prefill from an older request is ordered ahead of a fresh prefill from a newer one. The scheduler processes the queue and admits each prefill while it fits the remaining budget; the first that does not fit stops the scan.
When no prefill is waiting, the iteration is decode-only: the carried $\textit{deficit}$ is cleared to zero and the decode batch runs as is. Otherwise, whatever budget is left over is preserved as the next iteration's $\textit{deficit}$ and rolls into the next round.

Because dLLMs cannot chunk prefills, a single oversized prompt may exceed $\tau$ and might never be admitted into an idle system. We resolve this with an idle rule inside the admission loop: while the microbatch is still empty (no in-flight decode and no prefill admitted yet this iteration), the scheduler admits the head prefill even if it exceeds the budget. This is required for \textit{liveness} in the system.

Per-iteration stall-free batching is unattainable for dLLMs: an indivisible prefill must enter in a single iteration and stalls co-batched decodes by its full size. Due to this, prior dLLM serving has had to default to prefill-prioritizing schedulers ~\cite{fan2025taming} and pay the interference cost or be decode-prioritizing and suffer loss in throughput.
The deficit scheduler guarantees an amortized bound: over any busy period of $W$ iterations with prefills continuously waiting, admitted prefill tokens total at most\xspace$W\tau - \sum_{t=1}^{W} D_t$, so the average per-iteration prefill load is capped at $\tau$. In other words, the deficit construction substitutes temporal deferral for spatial chunking: rather than splitting one prefill across iterations, it defers whole prefills until enough budget accumulates, bounding the amortized prefill contribution to $\tau$. Temporal deferral is the only axis available once bidirectional attention rules out spatial chunking. Thus the scheduler achieves stall-free batching (in an amortized sense) \emph{without prefill chunking}, increasing the scope of stall-free continuous batching to a model class for which chunked prefill is structurally unavailable. The iteration token budget $\tau$ with deficit tracking provides a tunable knob over the queueing-delay versus prefill-decode interference trade-off for dLLMs, similar to AR serving. At a given request load, the budget should be high enough to clear prefills before they queue and low enough to shield decodes from interference.

\subsection{Hybrid Scheduling}\label{subsec:design-hybrid}

\begin{algorithm}[t]
\caption{Hybrid Scheduler}
\label{alg:hybrid}
\renewcommand{\algorithmicrequire}{\textbf{Input:}}
\begin{algorithmic}[1]
\Require request $r$; prefill pool $P$, colocated pool $C$; per-worker outstanding-prefill load $o_w$; overload threshold $\theta$; colocated deficit budget $\tau$
\State $w \gets \textsc{LeastLoaded}(P, r)$ \Comment{min $o_w$ under $\theta$, free KV}
\If{$w \neq \textbf{None}$}
    \State \textbf{dispatch} $r$ to $w$ as prefill-only; \textbf{return}
\EndIf
\State \Comment{prefill pool saturated or out of KV: overflow}
\State $w \gets \textsc{LeastLoaded}(C, r)$
\If{$w \neq \textbf{None}$ \textbf{and} colocated KV fits $r$ after pending decodes}
    \State \textbf{dispatch} $r$ to $w$ as colocated under $\tau$; \textbf{return}
\EndIf
\State \textbf{enqueue} $r$ on the FIFO pending queue \Comment{retry head-of-line later}
\end{algorithmic}
\end{algorithm}

A natural alternative to colocated serving is the prefill/decode disaggregated serving pioneered for AR LLMs~\cite{zhong2024distserve, patel2024splitwise, hu2024inference}: dedicate one pool of workers to prefill, another to decode, and transfer KV from the first to the second. Disaggregated serving eliminates prefill-decode interference by design. However, right sizing the prefill and decode pools is critical to improve goodput as shown by the design study ~\cite{mitra2025beyond}. To solve prefill under-provisioning which causes high queueing delays at prefill workers, \system{} pairs dedicated prefill workers with \emph{deficit token budget} based colocated workers in the decode role, rather than purely prefill-prioritizing or decode-prioritizing colocated workers. This reduces prefill-decode interference within the colocated workers while also absorbing queueing at the prefill workers.
A deployment provisions $N_P$ dedicated prefill workers and $N_C$ colocated workers. Under nominal load the central scheduler routes each request (and each block-boundary re-prefill) to the prefill pool, and the colocated pool serves the disaggregated decode role. When the prefill pool saturates, the scheduler \emph{overflows} the request to a colocated worker, which prefills and decodes the request locally with no cross-worker KV transfer. This seamlessly converts spare decode capacity into prefill capacity with prefill-decode interference controlled by deficit token budget.

The overflow trigger is token-based. The hybrid scheduler tracks for each worker, $o_w$, the outstanding prefill tokens the worker has been assigned but not yet processed. The prefill pool is considered overloaded only when \emph{every} non-draining prefill worker has $o_w \geq \theta$ for an operator-set threshold $\theta$ (in tokens).
If prefill overflow is detected, the scheduler dispatches the request to the least-loaded colocated worker that is still under $\theta$ and has free KV for it. Colocated KV is reserved first for the requests that already finished prefill and are waiting only on colocated memory to decode (the ``pending decodes'' of \Cref{alg:hybrid}).
If neither path can take the request, the scheduler holds it on a central FIFO pending queue and retries assignment as load or memory frees. 
Colocated workers run the deficit-budget scheduler with a tight per-iteration budget $\tau$ than prefill workers, since their primary role is decoding and overflow prefills should be admitted only opportunistically.

Hybrid scheduler interpolates between disaggregated and colocated serving. with $\theta{=}\infty$, overflow never fires, so the scheduler behaves as if purely disaggregated. With low $\theta$ and low $N_P/N_C$ ratio, hybrid exhibits colocated behavior as the colocated workers run both prefills and decodes. However, it cannot achieve fully colocated behavior because the prefill workers do not run any decodes. We deliberately omit migrating in-flight decodes onto dedicated workers because such decodes would suffer from uncontrolled interference from prefills in the prefill workers.

\subsection{\system Architecture}\label{subsec:architecture}
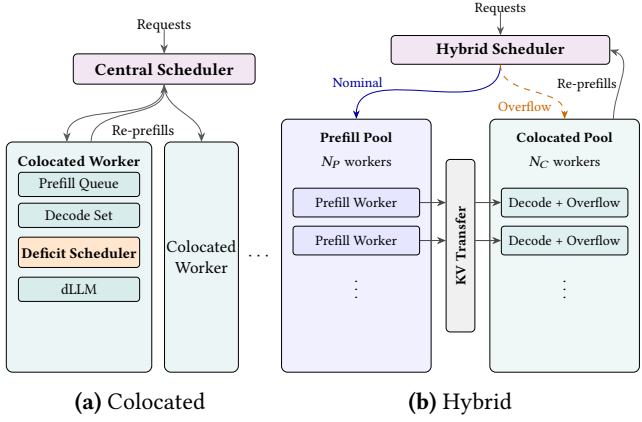
\begin{figure}[t]
    \centering
    \begin{subfigure}[b]{0.42\columnwidth}
\centering
\resizebox{\linewidth}{!}{%
\begin{tikzpicture}[
    font=\footnotesize,
    >={Stealth[length=1.5mm,width=1.2mm]},
    sched/.style={draw,rounded corners=2pt,fill=violet!10,
                  minimum width=30mm,minimum height=5mm,
                  inner sep=1pt,align=center,line width=0.3pt},
    worker/.style={draw,rounded corners=3pt,fill=teal!8,
                   minimum width=24mm,minimum height=38mm,
                   inner sep=1pt,line width=0.3pt},
    cworker/.style={draw,rounded corners=3pt,fill=teal!8,
                    minimum width=12mm,minimum height=38mm,
                    inner sep=1pt,align=center,line width=0.3pt},
    slot/.style={draw,rounded corners=1.5pt,fill=teal!15,
                 minimum width=20mm,minimum height=4mm,
                 inner sep=1pt,align=center,font=\scriptsize,line width=0.3pt},
    deficit/.style={draw,rounded corners=1.5pt,fill=orange!20,
                    minimum width=20mm,minimum height=5mm,
                    inner sep=1pt,align=center,font=\scriptsize\bfseries,line width=0.3pt},
    dots/.style={minimum width=4mm,minimum height=38mm,inner sep=1pt,align=center},
    arr/.style={->,line width=0.4pt,gray!70!black},
    lbl/.style={font=\scriptsize,inner sep=1pt}
]
    \node[sched] (sched) at (0,0) {\textbf{Central Scheduler}};

    \draw[arr] ([yshift=4mm]sched.north) -- (sched.north);
    \node[lbl,above=3mm of sched.north] {Requests};

    \node[worker,below=12mm of sched.west,xshift=1mm] (w1) {};
    \node[cworker,right=2mm of w1] (w2) {Colocated\\Worker};
    \node[dots,right=1mm of w2] (wd) {$\cdots$};

    \node[font=\scriptsize\bfseries,anchor=north] at ([yshift=-1mm]w1.north) {Colocated Worker};
    \node[slot,anchor=north] (q1) at ([yshift=-5mm]w1.north) {Prefill Queue};
    \node[slot,below=1mm of q1] (d1) {Decode Set};
    \node[deficit,below=1.5mm of d1] (s1) {Deficit Scheduler};
    \node[slot,below=1.5mm of s1] (k1) {dLLM};

    \draw[arr] (sched.south) to[out=-90,in=90] ([xshift=-2mm]w1.north);
    \draw[arr] (sched.south) to[out=-90,in=90] (w2.north);

    \draw[arr] ([xshift=2mm]w1.north) to[out=90,in=-90] (sched.south);
    \node[lbl,anchor=west] at ($(sched.south)!0.5!(w1.north)+(-2mm,-3mm)$) {Re-prefills};
\end{tikzpicture}%
}
\caption{Colocated}
\label{fig:colocated-arch}
\end{subfigure}\hfill
    \begin{subfigure}[b]{0.56\columnwidth}
\centering
\resizebox{\linewidth}{!}{%
\begin{tikzpicture}[
    font=\footnotesize,
    >={Stealth[length=1.5mm,width=1.2mm]},
    sched/.style={draw,rounded corners=2pt,fill=violet!10,
                  minimum width=38mm,minimum height=5mm,
                  inner sep=1pt,align=center,line width=0.3pt},
    pool/.style={draw,rounded corners=3pt,
                 minimum width=26mm,minimum height=44mm,
                 inner sep=1pt,line width=0.3pt},
    pcell/.style={draw,rounded corners=1.5pt,fill=blue!10,
                  minimum width=22mm,minimum height=5mm,
                  inner sep=1pt,align=center,font=\scriptsize,line width=0.3pt},
    ccell/.style={draw,rounded corners=1.5pt,fill=teal!12,
                  minimum width=22mm,minimum height=5mm,
                  inner sep=1pt,align=center,font=\scriptsize,line width=0.3pt},
    band/.style={draw,rounded corners=2pt,fill=gray!12,
                 minimum width=5mm,minimum height=30mm,
                 inner sep=1pt,align=center,font=\scriptsize\bfseries,line width=0.3pt},
    arrnom/.style={->,line width=0.5pt,blue!60!black},
    arrovf/.style={->,line width=0.5pt,orange!80!black,dashed},
    arrkv/.style={->,line width=0.4pt,gray!55!black},
    arr/.style={->,line width=0.4pt,gray!70!black},
    lbl/.style={font=\scriptsize,inner sep=1pt}
]
    \node[sched] (sched) at (0,0) {\textbf{Hybrid Scheduler}};

    \draw[arr] ([yshift=4mm]sched.north) -- (sched.north);
    \node[lbl,above=3mm of sched.north] {Requests};

    \node[pool,fill=blue!6,below=12mm of sched.west,xshift=-6mm] (pp) {};
    \node[band,right=2.5mm of pp] (kv) {\rotatebox{90}{KV Transfer}};
    \node[pool,fill=teal!6,right=2.5mm of kv] (cp) {};

    \node[font=\scriptsize\bfseries,anchor=north] (pph) at ([yshift=-1mm]pp.north) {Prefill Pool};
    \node[font=\scriptsize,anchor=north] at (pph.south) {$N_P$ workers};
    \node[font=\scriptsize\bfseries,anchor=north] (cph) at ([yshift=-1mm]cp.north) {Colocated Pool};
    \node[font=\scriptsize,anchor=north] at (cph.south) {$N_C$ workers};

    \node[pcell,anchor=north] (pw1) at ([yshift=-12mm]pp.north) {Prefill Worker};
    \node[pcell,below=1.5mm of pw1] (pw2) {Prefill Worker};
    \node[font=\scriptsize,below=1mm of pw2] (pdots) {$\vdots$};

    \node[ccell,anchor=north] (cw1) at ([yshift=-12mm]cp.north) {Decode + Overflow};
    \node[ccell,below=1.5mm of cw1] (cw2) {Decode + Overflow};
    \node[font=\scriptsize,below=1mm of cw2] (cdots) {$\vdots$};

    \draw[arrnom] (sched.south) to[out=-90,in=90] (pp.north);
    \draw[arrovf] (sched.south) to[out=-90,in=90] (cp.north);

    \node[lbl,blue!60!black,anchor=east] at ($(sched.south west)+(-1mm,-3mm)$) {Nominal};
    \node[lbl,orange!80!black,anchor=south west] at ($(cp.north)+(-12mm,1mm)$) {Overflow};

    \draw[arrkv] (pw1.east) -- (pw1.east -| kv.west);
    \draw[arrkv] (pw2.east) -- (pw2.east -| kv.west);
    \draw[arrkv] (kv.east |- cw1.west) -- (cw1.west);
    \draw[arrkv] (kv.east |- cw2.west) -- (cw2.west);

    \draw[arr] ($(cp.north)!0.65!(cp.north east)$) to[out=90,in=-20] (sched.east);
    \node[lbl,anchor=east] at ($(cp.north)!0.65!(cp.north east)+(1mm,6mm)$) {Re-prefills};
\end{tikzpicture}%
}
\caption{Hybrid}
\label{fig:hybrid-arch}
\end{subfigure}
    \caption{\textsc{\system} architectures. (a) Colocated: identical workers each run prefill and decode locally under the deficit-budget scheduler (\Cref{alg:deficit}). (b) Hybrid: dedicated prefill workers transfer KV to (primarily) decode-role colocated workers, with prefill overflow to colocated workers when all prefill workers exceed a load threshold $\theta$.}
    \label{fig:architectures}
\end{figure}

Based on the above design decisions, \system\ uses a two-level architecture. A central scheduler accepts incoming requests, tracks per-worker load, and routes each request (and each block-boundary re-prefill) to a worker. Workers run the model forward pass and apply the per-iteration batching policy locally. This split is needed because routing decisions need a global view of worker load and memory capacity usage across the cluster, while microbatch composition has to run on the worker's hot path between iterations and cannot afford a round-trip to a central process. \system{} instantiates this two-level architecture in two worker layouts.

\textbf{Colocated architecture (\Cref{fig:colocated-arch}).} Every worker is identical and runs prefill and decode locally. Each worker maintains a pending-prefill queue, an in-flight decode set, and admits work each iteration under the deficit-budget scheduler of \Cref{subsec:design-colocated}.

\textbf{Hybrid architecture (\Cref{fig:hybrid-arch}).} This is the instantiation of the hybrid scheduler described in \Cref{subsec:design-hybrid}.

A block-boundary re-prefill is not pinned to the worker that ran the previous block: dLLMs discard the per-request KV Cache at every block boundary and only the token IDs persist, so the central scheduler is free to redispatch a re-prefill to whichever worker is least loaded without paying any cross-worker KV transfer cost. This enables better load balancing across workers as the central scheduler can reassign requests at several points in their life-cycle.

\section{Implementation}
\label{sec:impl}

\system{} is implemented in roughly 16K lines of Python built on PyTorch and HuggingFace transformers. A central CPU-based scheduler coordinates a pool of GPU workers. Each worker can run as colocated, prefill-only, or decode-only, so the same system supports both architectures of Section~\ref{sec:design}. Control-plane RPCs use gRPC.
KV Cache transfers are streamed over \texttt{torch.distributed}/NCCL. The system uses PagedAttention~\cite{kwon2023efficient} for easy programmatic access to manage and transfer the KV cache. We also use FlashInfer's ~\cite{ye2025flashinfer} paged-KV attention kernels and CUDA Graphs as it significantly reduces iteration times.

\section{Evaluation}
\label{sec:eval}

We seek to answer the following questions in our evaluation:
\begin{itemize}
    \item \textbf{Q1:} How does \system compare against Fast-dLLM?
    \item \textbf{Q2:} How do colocated, disaggregated, and hybrid serving compare across different combinations of dLLM models and traces?
    \item \textbf{Q3:} Does deficit scheduling deliver amortized stall-free batching without prefill chunking, and how does the deficit token budget $\tau$ trade off throughput against inter token latency for colocated dLLM serving?
    \item \textbf{Q4:} How does the use of deficit token budget based hybrid scheduling impact its performance as the budget $\tau$ varies?
\end{itemize}

\subsection{Experiment Setup}
\label{subsec:eval-setup}
Experiments are run on a AWS p5.48xlarge node with 8 NVIDIA H100 GPUs each with 80\,GB  HBM, and connected via pairwise NVLink. We evaluate two open-source dLLMs, LLaDA-8B-Instruct~\cite{nie2026large} and Dream-7B-Instruct~\cite{ye2025dream}, both served in bfloat16. 
We use KV page size of 16 tokens (vLLM default), and block size of 32 (Fast-dLLM default). 
\begin{table}[t]
    \centering
    \setlength{\tabcolsep}{2.5pt}
    \begin{tabular}{lrrrrrr}
    \toprule
    & \multicolumn{3}{c}{Prompt Tokens} & \multicolumn{3}{c}{Output Tokens} \\
    \cmidrule(lr){2-4}\cmidrule(lr){5-7}
    Trace & Median & P99 & Std. & Median & P99 & Std. \\
    \midrule
    ShareGPT Chat & 795 & 2007 & 596 & 266 & 874 & 206 \\
    arXiv Summarization & 2827 & 3923 & 890 & 162 & 360 & 73 \\
    \bottomrule
    \end{tabular}
    \caption{Prompt-length and output-length distributions for the filtered ShareGPT and arXiv summarization traces.}
    \label{tab:eval-traces}
\end{table}

We use two traces: the ShareGPT ~\cite{sharegpt_vicuna_unfiltered} trace containing conversations with LLMs, and the arXiv summarization ~\cite{cohan2018discourse} trace containing research paper summarization tasks. We filter out requests having more than 4096 tokens and \Cref{tab:eval-traces} reports the prompt-length and decode-length distributions for the resulting traces. We append 1024 MASK tokens to the ShareGPT prompts, and 512 to the arXiv Summarization prompts to accommodate the longest outputs in these traces. 
Note that a request denoises one 32-position block at a time and advances only after a block resolves, so blocks beyond the committed EOS are never denoised or re-prefilled. A 266-token output processes ceil(266/32)=9 blocks, not the full L/32. The appended MASK tokens do however increase the sequence length processed in every prefill and re-prefill iterations.
We use a range of Poisson arrival rates, and report metrics from 4000-request runs per (configuration, QPS) point. Confidence-threshold based sampling with a threshold of 0.9 (Fast-dLLM default) is used to unmask these tokens representing a variable unmasking schedule.

We report end-to-end (E2E) latency, mean and P99, over a wide range of arrival rates. We deliberately avoid Time to First Token (TTFT) and Time Between Tokens (TBT) metrics from AR serving because these don't directly map to a dLLM request lifecycle. Due to presence of re-prefills, there is no single TTFT. Further, because bidirectional attention makes prefills indivisible, high P99 TBT is intrinsic to any colocated dLLM system and can only be bound in an amortized sense (\Cref{subsec:design-colocated}). To aid in understanding, we also report decode execution times, prefill execution times, and queueing delays at representative request load as needed.

\subsection{Comparison against Fast-dLLM}
\label{subsec:eval-fastdllm}
\begin{figure}[t]
    \centering
    \begin{subfigure}[t]{0.48\columnwidth}
        \centering
        \evalfig{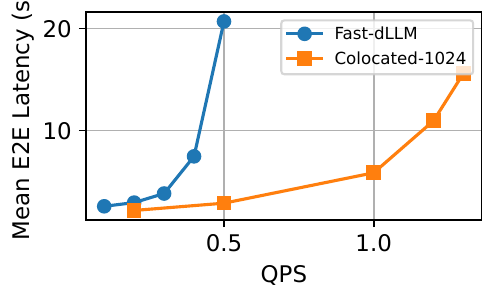}
        \caption{Mean, ShareGPT.}
        \label{fig:eval-fastdllm-mean-sharegpt}
    \end{subfigure}\hfill
    \begin{subfigure}[t]{0.48\columnwidth}
        \centering
        \evalfig{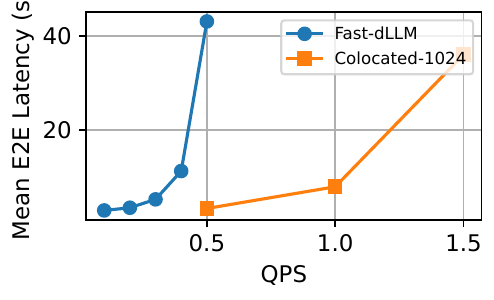}
        \caption{Mean, arXiv summarization.}
        \label{fig:eval-fastdllm-mean-arxiv}
    \end{subfigure}
    \begin{subfigure}[t]{0.48\columnwidth}
        \centering
        \evalfig{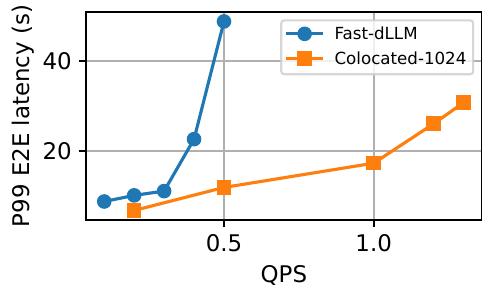}
        \caption{p99, ShareGPT.}
        \label{fig:eval-fastdllm-p99-sharegpt}
    \end{subfigure}\hfill
    \begin{subfigure}[t]{0.48\columnwidth}
        \centering
        \evalfig{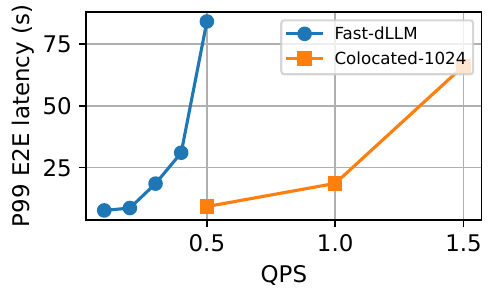}
        \caption{p99, arXiv summarization.}
        \label{fig:eval-fastdllm-p99-arxiv}
    \end{subfigure}
    \caption{Mean and p99 end-to-end latency under varying QPS for in-system Fast-dLLM and \system (Colocated with $\tau=1024$) running LLaDA-8B on a single H100 80GB GPU.}
    \label{fig:eval-fastdllm-latency}
\end{figure}

Fast-dLLM ~\cite{wu2025fast} ships as a inference script without an online server and without batching support for variable-length online traces, so it is run at batch size 1.
For a fair comparison we evaluate against an in-system version of Fast-dLLM: we take its model definition and use the same block KV caching and confidence-threshold sampling, but run it inside \system so it benefits from the optimizations such as CUDA Graph which reduce batch execution time by up to $2\times$.

On LLaDA-8B with the ShareGPT trace, Fast-dLLM's mean latency stays flat up to QPS\,0.3 and then increases sharply ($\sim$20.7\,s at QPS\,0.5). \system (Colocated, $\tau{=}1024$) instead sustains QPS\,1.0 at comparable mean and tail latency (\Cref{fig:eval-fastdllm-mean-sharegpt,fig:eval-fastdllm-p99-sharegpt}), a $\sim$3$\times$ throughput gap. On the arXiv trace Fast-dLLM knees by QPS\,0.3--0.4 while \system sustains QPS\,1.0 at comparable latency (\Cref{fig:eval-fastdllm-mean-arxiv,fig:eval-fastdllm-p99-arxiv}), a $\sim$2.5--3$\times$ gap. This massive difference in throughput under the same latency is because of insufficient GPU utilization by Fast-dLLM running without batching. Decode iterations at batch size 1 are highly inefficient as seen in \Cref{subsec:motiv-bigdecode-reprefill}.

We treat this comparison as a floor rather than a headline result. Because the in-system Fast-dLLM still serves one request at a time, the gap isolates the value of online batched serving, which \Cref{subsec:motiv-reprefills} motivates, not the value of deficit token budget based scheduler. The scheduler's contribution is evaluated against strong in-system baselines in \Cref{subsec:eval-colocated-vs-disagg}-~\Cref{subsec:eval-hybrid-scheduling}, where every configuration is online and batched.

\subsection{Colocated, Disaggregated, and Hybrid Serving}
\label{subsec:eval-colocated-vs-disagg}

\begin{figure*}[t]
    \centering
    \includegraphics[width=0.6\textwidth]{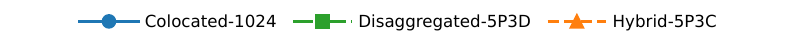}\\[-0.25em]
    \begin{subfigure}[t]{0.24\textwidth}
        \centering
        \evalfig[\linewidth]{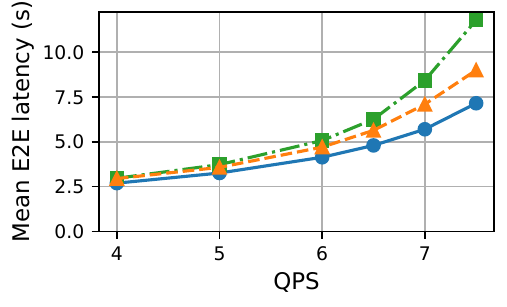}
        \caption{LLaDA-8B, ShareGPT, mean.}
        \label{fig:eval-cvsd-load-llada-sharegpt-mean}
    \end{subfigure}\hfill
    \begin{subfigure}[t]{0.24\textwidth}
        \centering
        \evalfig[\linewidth]{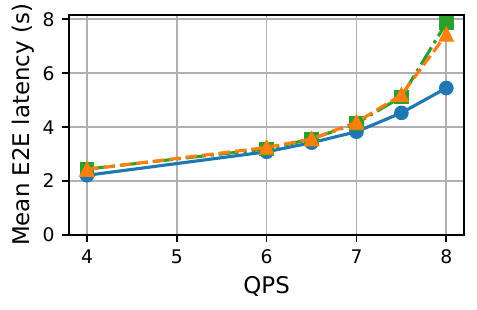}
        \caption{LLaDA-8B, arXiv, mean.}
        \label{fig:eval-cvsd-load-llada-arxiv-mean}
    \end{subfigure}\hfill
    \begin{subfigure}[t]{0.24\textwidth}
        \centering
        \evalfig[\linewidth]{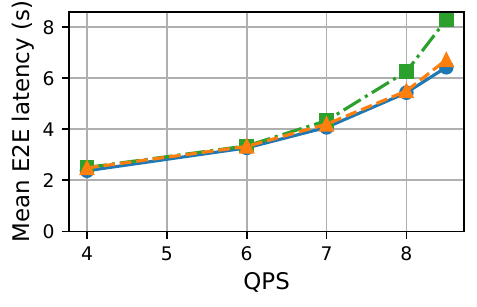}
        \caption{Dream-7B, ShareGPT, mean.}
        \label{fig:eval-cvsd-load-dream-sharegpt-mean}
    \end{subfigure}\hfill
    \begin{subfigure}[t]{0.24\textwidth}
        \centering
        \evalfig[\linewidth]{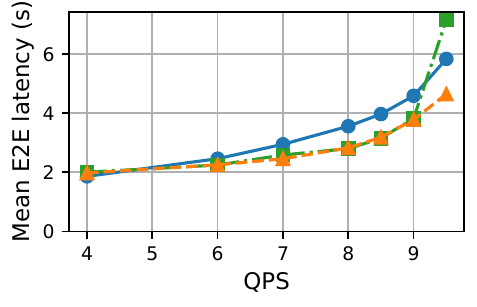}
        \caption{Dream-7B, arXiv, mean.}
        \label{fig:eval-cvsd-load-dream-arxiv-mean}
    \end{subfigure}

    \vspace{0.35em}
    \begin{subfigure}[t]{0.24\textwidth}
        \centering
        \evalfig[\linewidth]{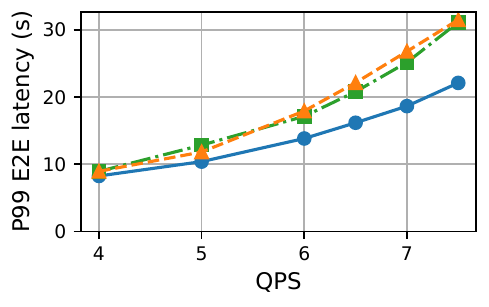}
        \caption{LLaDA-8B, ShareGPT, p99.}
        \label{fig:eval-cvsd-load-llada-sharegpt-p99}
    \end{subfigure}\hfill
    \begin{subfigure}[t]{0.24\textwidth}
        \centering
        \evalfig[\linewidth]{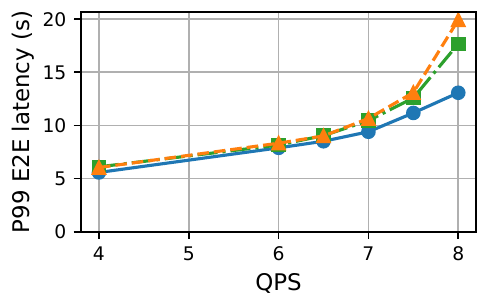}
        \caption{LLaDA-8B, arXiv, p99.}
        \label{fig:eval-cvsd-load-llada-arxiv-p99}
    \end{subfigure}\hfill
    \begin{subfigure}[t]{0.24\textwidth}
        \centering
        \evalfig[\linewidth]{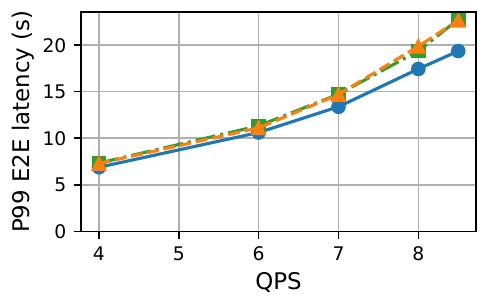}
        \caption{Dream-7B, ShareGPT, p99.}
        \label{fig:eval-cvsd-load-dream-sharegpt-p99}
    \end{subfigure}\hfill
    \begin{subfigure}[t]{0.24\textwidth}
        \centering
        \evalfig[\linewidth]{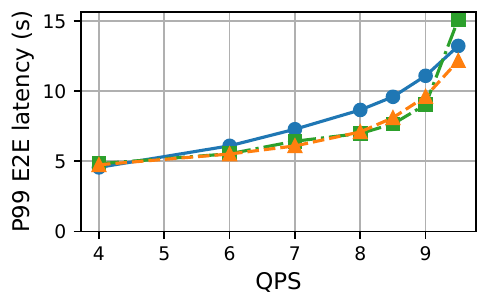}
        \caption{Dream-7B, arXiv, p99.}
        \label{fig:eval-cvsd-load-dream-arxiv-p99}
    \end{subfigure}
    \caption{End-to-end latency vs.\ QPS for colocated, static-split disaggregated (5P3D), and hybrid (5P3C). Top row mean, bottom row p99; columns are (model, trace) pairs. Colocated and hybrid degrade gracefully while the static disaggregated typically saturates first.}
    \label{fig:eval-cvsd-load}
\end{figure*}

\begin{figure}[t]
    \centering
    \includegraphics[width=0.8\columnwidth]{figures/cvsd-plots/c_vs_d_legend.pdf}\\[-0.25em]
    \begin{subfigure}[t]{0.49\columnwidth}
        \centering
        \evalfig[\linewidth]{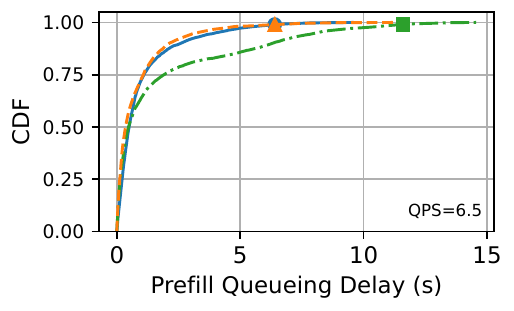}
        \caption{LLaDA-8B ShareGPT.}
        \label{fig:eval-cvsd-cdf-llada-sharegpt-prefill}
    \end{subfigure}\hfill
    \begin{subfigure}[t]{0.49\columnwidth}
        \centering
        \evalfig[\linewidth]{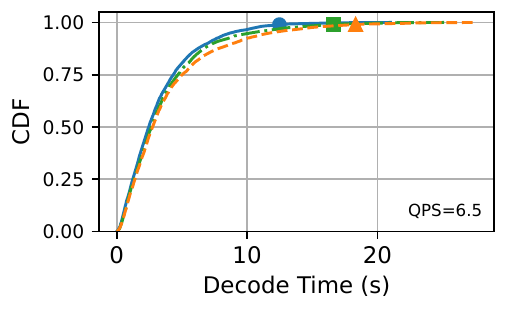}
        \caption{LLaDA-8B ShareGPT.}
        \label{fig:eval-cvsd-cdf-llada-sharegpt-decode}
    \end{subfigure}

    \vspace{0.35em}
    \begin{subfigure}[t]{0.49\columnwidth}
        \centering
        \evalfig[\linewidth]{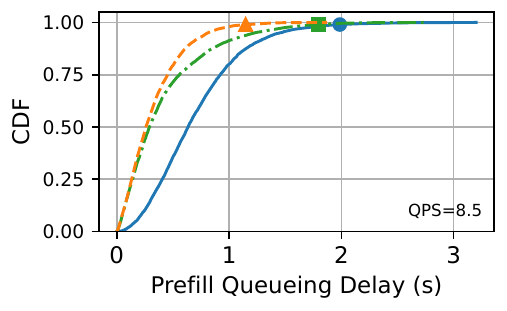}
        \caption{Dream-7B arXiv.}
        \label{fig:eval-cvsd-cdf-dream-arxiv-prefill}
    \end{subfigure}\hfill
    \begin{subfigure}[t]{0.49\columnwidth}
        \centering
        \evalfig[\linewidth]{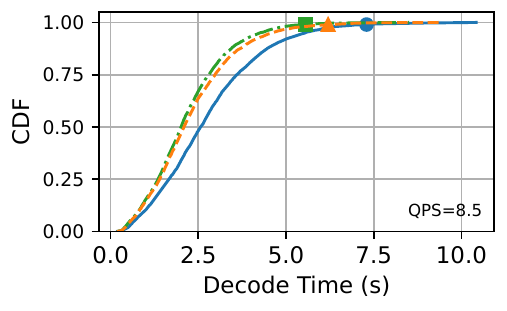}
        \caption{Dream-7B arXiv.}
        \label{fig:eval-cvsd-cdf-dream-arxiv-decode}
    \end{subfigure}
    \caption{Per-request CDFs of prefill queueing delay (left) and request decode time (right), at moderate request load (LLaDA-8B ShareGPT 6.5\,QPS, Dream-7B arXiv 8.5\,QPS). p99 values are highlighted.
    }
    \label{fig:eval-cvsd-cdf}
\end{figure}

We next ask whether dLLM serving should use separate pools of GPUs to prefill and decode, or let every GPU serve both phases.
We compare three ways \textit{(colocated, disaggregated and hybrid)} to configure an 8-GPU node, picking a strong configuration of each.
The \emph{colocated} mode runs all eight GPUs as colocated workers with a 1024-token iteration budget which strikes a balance between reducing prefill-decode interference and throughput as will be shown in \Cref{subsec:eval-deficit}.
The \emph{disaggregated} configuration has fixed prefill and decode pools; we use a 5P3D split (five prefill workers, three decode workers), the best static split for these workloads, since 4P4D is even more prefill-bound and 6P2D runs the decode pool into its KV Cache capacity limit. \emph{Hybrid 5P3C} (\Cref{subsec:design-hybrid}) reuses the same split as disaggregated, but its three decode-role workers are colocated workers. We run hybrid with overflow threshold $\theta{=}8\mathrm{k}$ and a tight deficit token budget $\tau{=}1024$. 
This specific hybrid configuration is used to show the gains which can be achieved by simply replacing decode workers with tight budget colocated workers in static disaggregated deployments.

\textbf{\textit{Overall trends.}} \Cref{fig:eval-cvsd-load} reports mean and p99 end-to-end latency across a range of QPS for four (model, trace) pairs. At low load the three configurations are effectively tied; as load rises they diverge. Disaggregated-5P3D saturates first in every pair: on LLaDA-8B ShareGPT its mean latency reaches 11.8\,s at QPS\,7.5 against 7.1\,s for colocated and 9.0\,s for hybrid. Colocated has the lowest mean and p99 in three of the four pairs, the exception being the prefill-heavy Dream-7B arXiv, where disaggregated and hybrid stay ahead of colocated across almost the whole sweep (at QPS\,7 disaggregated is about 12\% lower than colocated at both mean and p99, hybrid 17\% and 16\%) until the static disaggregated finally saturates at the highest QPS. Hybrid stays close to the better of the two: it follows colocated through the loaded region while capturing the disaggregated advantage on Dream-7B arXiv.

These trends come from two metrics that distinguish the schedulers: prefill queueing delay and request decode time, which are the largest contributors to end-to-end latency. \Cref{fig:eval-cvsd-cdf} shows the per-request CDFs of both at a representative QPS for two illustrative pairs: the decode-heavy LLaDA-8B ShareGPT and the prefill-heavy Dream-7B arXiv, which bracket the behavior of the remaining two pairs. This is because the arXiv trace has longer prompts than ShareGPT (\Cref{tab:eval-traces}) and Grouped Query Attention (GQA)~\cite{ainslie2023gqa} in Dream-7B which shrinks the KV Cache footprint and hence the cost of memory-bandwidth bound decodes ~\cite{agrawal2024vidur} compared to LLaDA-8B which uses MHA.

\textbf{\textit{Disaggregated Serving.}} Disaggregated dedicates fixed pools to each phase (five prefill, three decode here) to remove prefill-decode interference and hence keep decode times low. The cost is a resource partitioning problem: because the split is static, instantaneous load fluctuations on the prefill side, from new request arrivals and data-dependent re-prefills, queue up behind the five prefill workers. On LLaDA-8B ShareGPT (\Cref{fig:eval-cvsd-cdf-llada-sharegpt-prefill}) this prefill queueing delay reaches 6.3\,s p90 and 11.6\,s p99, against 2.6\,s and 6.4\,s for colocated. The same partitioning problem surfaces on Dream-7B arXiv too, but only at the highest load, where disaggregated's prefill pool saturates at QPS\,9.5 and its mean latency jumps to 7.2\,s against colocated's 5.8\,s (\Cref{fig:eval-cvsd-load-dream-arxiv-mean}). Static partitioning doesn't always face imbalance though. On Dream-7B arXiv, disaggregated has lesser prefill queueing than colocated as the model, workload, and partitioning align \Cref{fig:eval-cvsd-cdf-dream-arxiv-prefill}.

\textbf{\textit{Colocated Serving.}} Colocated lets every worker run both phases, so it intrinsically absorbs load fluctuations, avoiding the prefill queue. But a small token budget nevertheless would result in high prefill queueing delay. This is seen in LLaDA-8B ShareGPT (\Cref{fig:eval-cvsd-cdf-llada-sharegpt-prefill}), where colocated has the lowest prefill queueing delay across the CDF. For Dream-7B arXiv (\Cref{fig:eval-cvsd-cdf-dream-arxiv-prefill}) , $\tau{=}1024$ the budget is not enough for this trace so prefill queueing delay is highest.
On the decode side, colocated spreads decodes across all eight workers, shrinking decode batch sizes and thus decode batch times. For autoregressive models this would under utilize the GPU compute, but dLLM decodes are heavy enough that a handful of them saturate GPU utilization (\Cref{subsec:motiv-bigdecode-reprefill}). Decodes also experience interference in colocated which is controlled by the token budget.
The net decode cost is model and workload dependent. 
On prefill-heavy Dream-7B arXiv (\Cref{fig:eval-cvsd-cdf-dream-arxiv-decode}), with long prompts and few decodes, interference dominates and disaggregated keeps the lowest decode time (5.6\,s p99 against colocated's 7.3\,s). On the decode-heavy LLaDA-8B ShareGPT trace (\Cref{fig:eval-cvsd-cdf-llada-sharegpt-decode}) spreading roughly offsets interference and the two are close at the median (2.5\,s versus 2.7\,s).

\textbf{\textit{Hybrid Serving.}} Hybrid-5P3C replaces the dedicated decode workers with tight-budget colocated workers, so when its prefill pool saturates the overflow path spreads prefills across all GPUs. It therefore drains prefill queueing in every case: on LLaDA-8B ShareGPT it matches colocated (6.4\,s p99, well below disaggregated's 11.6\,s), and on Dream-7B arXiv it is the lowest of the three (1.1\,s p99). The cost is interference between overflow prefills and the decodes on those colocated workers, which lifts hybrid decode time above disaggregated (18.3\,s versus 16.6\,s p99 on ShareGPT, 6.2\,s versus 5.6\,s on arXiv). Overall, hybrid matches or improves the mean end-to-end latency over pure disaggregated in all cases and tracks its tail closely.

Finally, KV Cache transfer in disaggregated and hybrid is a negligible overhead with p99 value at 384\,ms on LLaDA-8B ShareGPT and 192\,ms on Dream-7B arXiv against multi-second decodes due to fast intra-node links. One can argue that inter-node bandwidths are lower, but this can offset by layerwise KV streaming and transfer-decode overlap ~\cite{sun2024llumnix,strati2024d,patel2024splitwise}, which we do not employ. Prior work also indicates that KV transfer is not the primary bottleneck in disaggregated serving ~\cite{mitra2025beyond}.

\subsection{Colocated Scheduling}
\label{subsec:eval-deficit}

\begin{figure*}[t]
    \centering
    \includegraphics[width=0.6\textwidth]{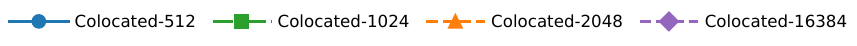}\\[-0.25em]
    \begin{subfigure}[t]{0.24\textwidth}
        \centering
        \evalfig[\linewidth]{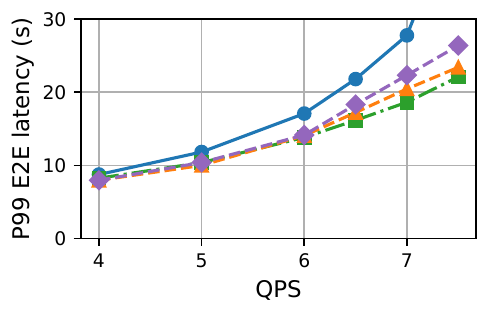}
        \caption{LLaDA-8B, ShareGPT}
        \label{fig:eval-deficit-llada-sharegpt-load}
    \end{subfigure}\hfill
    \begin{subfigure}[t]{0.24\textwidth}
        \centering
        \evalfig[\linewidth]{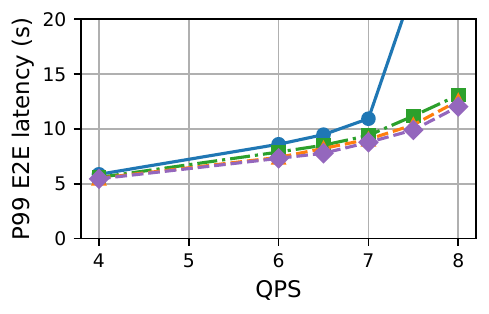}
        \caption{LLaDA-8B, arXiv}
        \label{fig:eval-deficit-llada-arxiv-load}
    \end{subfigure}\hfill
    \begin{subfigure}[t]{0.24\textwidth}
        \centering
        \evalfig[\linewidth]{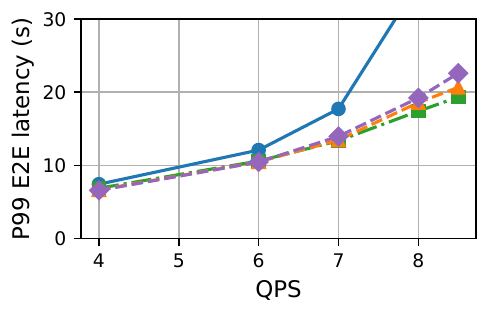}
        \caption{Dream-7B, ShareGPT}
        \label{fig:eval-deficit-dream-sharegpt-load}
    \end{subfigure}\hfill
    \begin{subfigure}[t]{0.24\textwidth}
        \centering
        \evalfig[\linewidth]{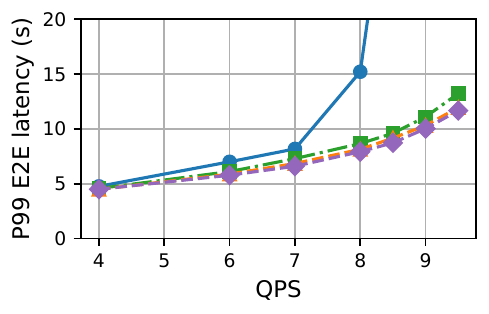}
        \caption{Dream-7B, arXiv}
        \label{fig:eval-deficit-dream-arxiv-load}
    \end{subfigure}

    \vspace{0.35em}
    \includegraphics[width=0.6\textwidth]{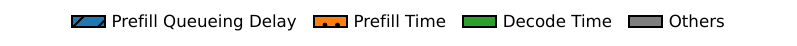}\\[-0.25em]
    \begin{subfigure}[t]{0.24\textwidth}
        \centering
        \evalfig[\linewidth]{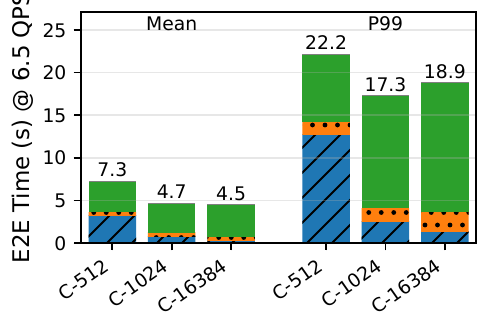}
        \caption{LLaDA-8B, ShareGPT}
        \label{fig:eval-deficit-llada-sharegpt-breakdown}
    \end{subfigure}\hfill
    \begin{subfigure}[t]{0.24\textwidth}
        \centering
        \evalfig[\linewidth]{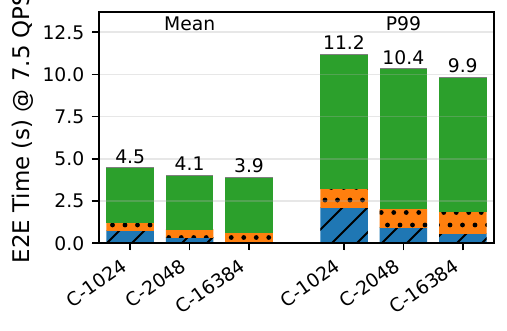}
        \caption{LLaDA-8B, arXiv}
        \label{fig:eval-deficit-llada-arxiv-breakdown}
    \end{subfigure}\hfill
    \begin{subfigure}[t]{0.24\textwidth}
        \centering
        \evalfig[\linewidth]{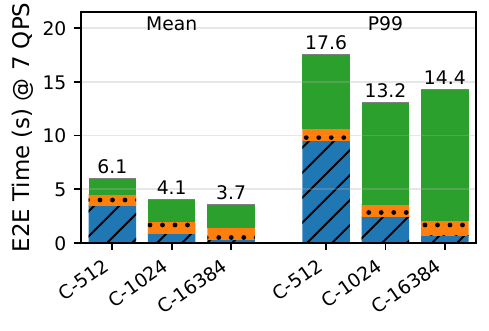}
        \caption{Dream-7B, ShareGPT}
        \label{fig:eval-deficit-dream-sharegpt-breakdown}
    \end{subfigure}\hfill
    \begin{subfigure}[t]{0.24\textwidth}
        \centering
        \evalfig[\linewidth]{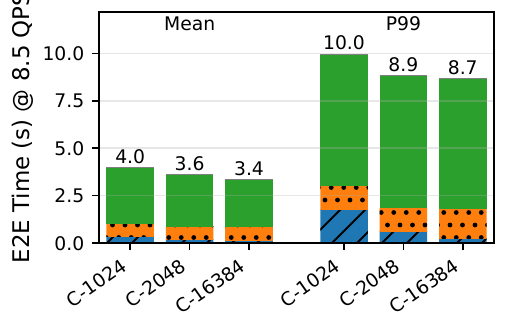}
        \caption{Dream-7B, arXiv}
        \label{fig:eval-deficit-dream-arxiv-breakdown}
    \end{subfigure}
    \caption{Deficit token-budget sweep $\tau \in \{512, 1024, 2048, 16384\}$ for colocated serving. The top row reports a p99 end-to-end latency load curve (vs.\ QPS) for each (model, trace) pair across all four budgets. The bottom row decomposes a single mean and tail (p99) request, taken at one fixed QPS in that pair's stable regime, into prefill queueing delay, prefill time, decode time, and other. Columns are (model, trace) pairs.}
    \label{fig:eval-deficit}
\end{figure*}

We demonstrate that deficit token budget based colocated scheduling reduces request decode times by reducing prefill-decode interference, and that the budget trades tail latency against the lower percentiles: a lower budget reduces the tail at the cost of raising the median, and vice versa.
To do this we evaluate a suite of iteration token budgets $\tau \in \{512, 1024, 2048, 16384\}$. $\tau{=}512$ is the smallest per-iteration budget we test where most prefills are deferred for several iterations until carryover budget accumulates. As $\tau$ grows the scheduler admits prefills more eagerly, and at $\tau{=}16384$ the budget is high enough that any prefill is processed in the immediate next iteration of its arrival (when memory capacity is also available), so the scheduler degenerates to the prefill-prioritizing behavior of Orca ~\cite{yu2022orca} and vLLM~\cite{kwon2023efficient}. This prefill-prioritizing budget is an in-system realization of dLLM-Serve's ~\cite{fan2025taming} refresh-reuse policy, reproduced under identical model, trace, and hardware.

\Cref{fig:eval-deficit} reports, for every (model, trace) pair, a p99 end-to-end latency curve wrt. load (top row) and the end-to-end latency breakdown of a mean and a tail (p99) request at one specific QPS (bottom row), representative of the scheduler behavior. The breakdown picks a single request id within a small band around every budget's own e2e latency (one at p50, and another at p99), widening the band until one id qualifies in all of them. Each iteration has a prefill budget according the value of $\tau$, so as QPS rises the arrival rate of fresh prefills and re-prefills outpaces what the scheduler can admit. Requests then queue until enough deficit accumulates, and queueing delay grows super-linearly. With $\tau{=}512$ each iteration carries minimal residual capacity, so its load curve develops this knee well before others. For example, on LLaDA-8B ShareGPT its p99 climbs from 27.8\,s at QPS\,7 to 45.6\,s at QPS\,7.5, reaching close to $2\times$ the best budget by QPS\,7.5 (45.6\,s versus 22.1\,s for $\tau{=}1024$).

In the ShareGPT trace, $\tau{=}1024$ holds the lowest p99 across a broad range on both models (between QPS\,6-7.5 in LLaDA-8B and QPS\,7-8.5 in Dream-7B), keeping a margin of up to roughly 10\% over the next best $\tau{=}2048$.  The breakdown figures (\Cref{fig:eval-deficit-llada-sharegpt-breakdown,fig:eval-deficit-dream-sharegpt-breakdown}) show why, and reveal that the budget trades the tail against the lower percentiles. On LLaDA-8B ShareGPT, raising $\tau$ from 1024 to 16384 nearly halves the tail request's queueing delay (2.5\,s to 1.4\,s) but raises its decode time (13.1\,s to 15.1\,s) as more prefills interfere with concurrent decodes, so the tail end-to-end climbs from 17.3\,s to 18.9\,s. Dream-7B ShareGPT shows the same pattern: the tail decode time climbs from 9.5\,s to 12.3\,s and end-to-end from 13.2\,s to 14.4\,s. The mean request moves the opposite way: at $\tau{=}16384$ queueing is already negligible, so the lower budget buys it nothing and only adds interference, leaving the high budget with the lower mean (4.5\,s against 4.7\,s on LLaDA-8B, 3.7\,s against 4.1\,s on Dream-7B). The mechanism behind this split is that a high budget eagerly admits prefills as soon as they arrive, sharply stalling the decodes being processed close to prefill arrivals. A lower budget instead defers prefills and spreads processing across many iterations, thinning the same interference over many more decodes: this reduces the worst-case stall (a lower tail) but increases the end-to-end latency at other percentiles i.e. a higher mean. Pushing the budget too low overshoots, as deferred prefills then queue for a long time: at $\tau{=}512$ the LLaDA-8B tail request waits 12.8\,s in the queue, pushing its end-to-end to 22.2\,s, above both larger budgets, so the tail optimum is intermediate. Note that as the request load continues to increase, $\tau{=}1024$ will reach its knee point before higher budgets as requests will queue faster than the prefill processing capacity $\tau{=}1024$ allows. Post that point, queueing delay in $\tau{=}1024$ will grow exponentially, giving way to $\tau{=}2048$ as having the lowest latency, and so on.

In the arXiv trace, we find that the highest budget $\tau{=}16384$ performs the best, although by a vanishing margin compared to the next lower budgets $\tau{=}2048$ and $\tau{=}1024$. The smallest budget $\tau{=}512$ saturates first: on
LLaDA-8B arXiv \cref{fig:eval-deficit-llada-arxiv-load} $\tau{=}512$ saturates at QPS\,7 (10.9\,s, climbing to 49.7\,s at QPS\,8), while the larger budgets remain near 12-13\,s at QPS\,8. Thus for this trace, there isn't gains to be had in E2E times by adjusting the budget because of the prefill heavy nature of the arxiv trace (\Cref{subsec:eval-setup}).
On a prefill-heavy trace the dominant requirement is to clear prefills before they queue, which a high budget allows. Only in a decode-heavy trace there is enough decode volume to protect from interference, so there exists a specific budget such as $\tau{=}1024$ for ShareGPT, which defers prefills and provides gains. The E2E time breakdown plots \Cref{fig:eval-deficit-llada-arxiv-breakdown} and \Cref{fig:eval-deficit-dream-arxiv-breakdown} corroborate this.

\subsection{Hybrid Scheduling}
\label{subsec:eval-hybrid-scheduling}

\begin{figure}[t]
    \centering
    \begin{subfigure}[t]{0.48\linewidth}
        \centering
        \evalfig[\linewidth]{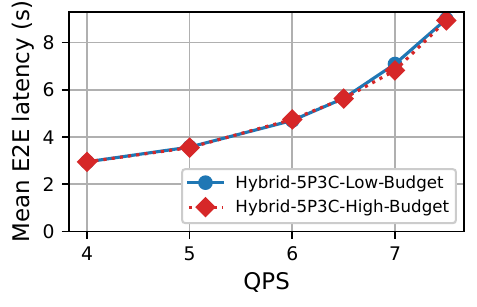}
        \caption{5P3C, mean e2e latency.}
        \label{fig:eval-hybrid-budget-5p3c-mean}
    \end{subfigure}\hfill
    \begin{subfigure}[t]{0.48\linewidth}
        \centering
        \evalfig[\linewidth]{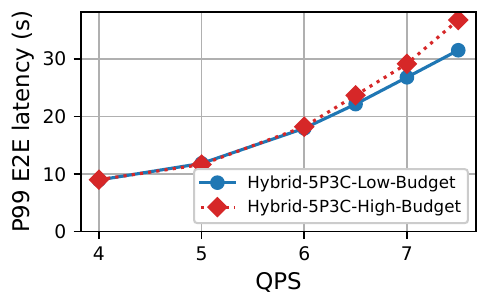}
        \caption{5P3C, p99 e2e latency.}
        \label{fig:eval-hybrid-budget-5p3c-p99}
    \end{subfigure}

    \vspace{0.4em}
    \begin{subfigure}[t]{0.48\linewidth}
        \centering
        \evalfig[\linewidth]{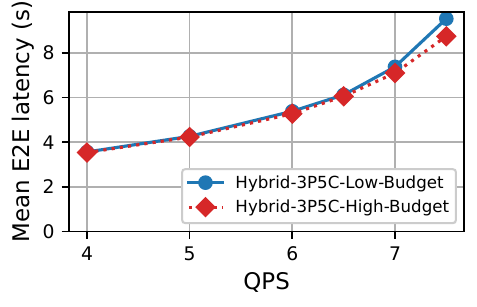}
        \caption{3P5C, mean e2e latency.}
        \label{fig:eval-hybrid-budget-3p5c-mean}
    \end{subfigure}\hfill
    \begin{subfigure}[t]{0.48\linewidth}
        \centering
        \evalfig[\linewidth]{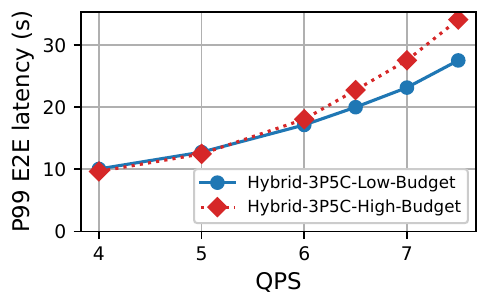}
        \caption{3P5C, p99 e2e latency.}
        \label{fig:eval-hybrid-budget-3p5c-p99}
    \end{subfigure}
    \caption{Sensitivity of hybrid scheduling to the deficit token budget $\tau$ on LLaDA-8B with the ShareGPT trace, at the 5P3C (top row) and 3P5C (bottom row) ratios; each row reports mean (left) and p99 (right) end-to-end latency vs.\ QPS, with a low ($\tau{=}1024$) and a high ($\tau{=}4096$) budget. 
    }
    \label{fig:eval-hybrid-budget}
\end{figure}

We seek to understand how the deficit token budget $\tau$ impacts hybrid scheduling.
\Cref{fig:eval-hybrid-budget} shows load plots on LLaDA-8B with the ShareGPT trace at the 5P3C and 3P5C prefill:colocated worker ratios, comparing a low ($\tau{=}1024$) against a high ($\tau{=}4096$) colocated budget. We use an overflow threshold of $\theta{=}8192$ as it performs well in a sweep of hyperparamter combinations in hybrid serving. 
At the 5P3C configuration (\Cref{fig:eval-hybrid-budget-5p3c-mean,fig:eval-hybrid-budget-5p3c-p99}), the lower budget ($\tau{=}1024$) reduces the p99 end-to-end latency at high load, about 8\% at QPS\,7 (26.8\,s against 29.1\,s) and 14\% at QPS\,7.5 (31.5\,s against 36.7\,s). This is because a lower budget is more effective in keeping overflow prefills from stalling decodes already in flight on the colocated workers.
Similarly in the 3P5C configuration (\Cref{fig:eval-hybrid-budget-3p5c-p99}), the low budget again reduces the p99 latency roughly 5\% at QPS\,6 rising to 16--19\% at high load (27.5\,s against 34.1\,s at QPS\,7.5). In both the cases, the impact is visible at high load because the prefill overflow condition triggers significantly more frequently at high load.

The reduction in p99 latencies by using low budgets is not free. Processing overflow prefills slowly via a low budget means the colocated workers spare less prefill capacity to the system, so prefills queue longer in the dedicated prefill pool and the mean end-to-end latency rises. This presents itself in the 3P5C configuration (\Cref{fig:eval-hybrid-budget-3p5c-mean}), where the prefill pool is under-provisioned to begin with. Here the mean rises, from about 2\% higher at QPS\,6 to 9\% at QPS\,7.5 (9.5\,s against 8.7\,s) as the three-worker prefill pool backs up. For the 5P3C configuration (\Cref{fig:eval-hybrid-budget-5p3c-mean}), the mean end-to-end latency is however essentially unchanged (within a few percent) because the five-worker prefill pool is large enough to absorb the prefill work the colocated workers no longer take on.

\section{Related Work}
\label{sec:related}

\textbf{dLLM architectures.} The dLLM design space is large and continuously expanding ~\cite{li2025survey}, with no single serving solution covering it. One strain of models restores causal structure to reduce inference footprint by exact KV caching. Block Diffusion ~\cite{arriola2025block} attends causally across blocks and bidirectionally within them, leading to models such as SDAR~\cite{cheng2026sdar}, Fast-dLLM v2 \cite{wu2025fast2}, LLaDA2-100B ~\cite{bie2025llada2}, and IDLM ~\cite{yu2026introspective}. Because these models behave exactly like an AR model at the block level, LLM serving engines such as SGLang ~\cite{zheng2024sglang} has added support for these dLLMs, as competitors to speculative decoding ~\cite{leviathan2023fast, li2024eagle, liu2024deepseek}. Another strain of models keep attention bidirectional, with LLaDA-8B ~\cite{nie2026large} and Dream-7B ~\cite{ye2025dream, xie2025dream} as widely used instances. An active line of work continues to advance dLLMs through new model architectures ~\cite{rout2026anchored, hersche2026softmasked, fu2026nemotron, sahoo2026esoteric}, sampling strategies ~\cite{NEURIPS2025_4c7d31b2,bansal2025enabling,zhao2026d1}, and scaling laws ~\cite{nie2025scaling,gulrajani2023likelihood,von2025scaling}.

We pick two popular, representative scenarios, LLaDA-8B and Dream-7B with Fast-dLLM blockwise caching, to make a methodological point: approximate KV caching induces a cyclic prefill/decode structure in dLLM inference and makes it amenable to the AR serving stack. We then identify the critical differences, block-sized decodes, recurring prefills, and the absence of chunked prefill, and adapt the stack to them. Because this cyclic structure is shared by any cached dLLM, including block-causal models that simply re-prefill less often, the same mechanisms and the same partitioning-versus-interference analysis carry across the class.

\textbf{dLLM serving engines.} To our knowledge, dInfer~\cite{ma2025dinfer} and dLLM-Serve~\cite{fan2025taming} are the only other serving engines built for LLaDA-8B and Dream-7B. dInfer contributes parallel decoding strategies, approximate K/V caching, and kernel optimizations, evaluated at batch size 1. Its optimizations target the decoding algorithm and are orthogonal to Sangam's scheduling, so we do not compare directly. dLLM-Serve likewise centers on algorithmic techniques such as decomposing transient logit peaks and sparsifying attention storage and proposes a vLLM-style prefill-prioritizing policy for prefill-decode (termed as "refresh-reuse") scheduling. We reproduce this prefill-prioritizing policy as the $\tau=16384$ colocated configuration in ~\Cref{subsec:eval-deficit}, rather than comparing against the dLLM-Serve codebase, which is tightly coupled with its other algorithmic proposals, and show it incurs the prefill-decode interference that motivates this work.

\section{Conclusion}

Approximate KV caching gives diffusion language models a repeated prefill/decode structure, making the AR serving stack a useful starting point for dLLM serving. However, cached dLLMs violate several assumptions behind that stack: decodes operate on response blocks rather than single tokens, prefills recur at block boundaries or cache-refresh points, and bidirectional attention prevents chunked prefill. \system{} addresses these differences with a deficit token-budget scheduler that provides amortized stall-free colocated serving for indivisible dLLM prefills, and with a serving architecture that supports colocated, static disaggregated, and hybrid prefill/decode execution under one implementation. Our evaluation on LLaDA-8B and Dream-7B shows that the effectiveness of colocated, disaggregated, and hybrid serving is governed by two fundamental factors, static prefill/decode partitioning and prefill-decode interference, and that hybrid scheduling is a simple augmentation that preserves the strengths of static disaggregated deployments while improving them on decode-heavy workloads. \system{} shows that a disciplined reuse of the AR stack is the right foundation for serving the growing class of cached dLLMs.

\bibliographystyle{ACM-Reference-Format}
\bibliography{references}

\end{document}